\documentclass[pdflatex, sn-mathphys]{sn-jnl}% Math and Physical Sciences Numbered Reference Style 

\usepackage{graphicx}%
\usepackage{multirow}%
\usepackage{amsmath,amssymb,amsfonts}%
\usepackage{amsthm}%
\usepackage{mathrsfs}%
\usepackage[title]{appendix}%
\usepackage{xcolor}%
\usepackage{textcomp}%
\usepackage{manyfoot}%
\usepackage{booktabs}%
\usepackage{algorithm}%
\usepackage{algorithmicx}%
\usepackage{algpseudocode}%
\usepackage{listings}%
\usepackage{multicol}
\usepackage{footmisc}
\usepackage{sidecap}
\usepackage{comment} 
\usepackage{upgreek}
\usepackage{xspace}

\usepackage{natbib}
\usepackage{subcaption}

\usepackage{bm}
\usepackage{bbm} 
\usepackage[overload]{empheq} % For braced-style systems of equations.
\usepackage{fix-cm} % To override original LaTeX restrictions on sizes

\usepackage{enumitem}
\usepackage{cleveref}
\usepackage{nicefrac}
\usepackage{tabularx}
\usepackage{anyfontsize}

\usepackage{caption}  % per biblio

\usepackage{glossaries}       % For glossary management

\newacronym{mqsrpde}{MQSR}{Multiple Quantile Spatial Regression with Partial Differential Equation regularization}
\newacronym{gcv}{GCV}{Generalized Cross-Validation}
\newacronym{fem}{FEM}{Finite Element Method}
\newacronym{fpirls}{FPIRLS}{Functional Penalized Iterative Reweighted Least Squares}
\newacronym{arpa}{ARPA}{Agenzia Regionale per la Protezione dell'Ambiente}
\newacronym{mm}{MM}{Maximization-Minimization}
\newacronym{pde}{PDE}{Partial Differential Equation}

\crefname{equation}{Equation}{Equations}
\crefname{figure}{Figure}{Figures}
\crefname{section}{Section}{Sections}
\crefname{chapter}{Chapter}{Chapters}
\crefname{appendix}{Appendix}{Appendices}

\usepackage{url}

\usepackage{tikz}
\usepackage{array}

\newcommand{\PMten}{PM\textsubscript{10}\xspace}

\newcommand{\authors}[1]{\textcolor{black}{#1}}

\begin{document}

\title[Article Title]{Three Distributional Approaches for PM$_{10}$ Assessment in Northern Italy}


\author[1]{\fnm{Marco F.} \sur{De Sanctis}}
\author[1,2]{\fnm{Andrea} \sur{Gilardi}}
\author[1]{\fnm{Giacomo} \sur{Milan}}
\author[1]{\fnm{Laura M.} \sur{Sangalli}}
\author[1]{\fnm{Francesca} \sur{Ieva}}
\author[1]{\fnm{Piercesare} \sur{Secchi}}

\affil[1]{\orgdiv{MOX, Dipartimento di Matematica}, \orgname{Politecnico di Milano}, \orgaddress{\street{Piazza Leonardo Da Vinci 32}, \city{Milano}, \postcode{20133}, 
\state{Italy}}
}

\affil[2]{\orgdiv{Dipartimento di Economia, Metodi Quantitativi e Strategie d'Impresa (DEMS)}, \orgname{Università degli Studi di Milano - Bicocca}, \orgaddress{\street{Piazza dell'Ateneo Nuovo 1}, \city{Milano}, \postcode{20126}, 
\state{Italy}}
}

\abstract{Air quality assessment often relies on multiple distributional summaries, including expected concentrations, quantiles, and regulatory exceedance probabilities. Standard geostatistical approaches typically model these quantities separately. Here, we reconstruct the entire distribution of PM\textsubscript{10} concentrations through three representations of spatially varying pollutant distributions. The proposed representations capture pollutant distributions at increasing levels of detail: as a two-part composition, through multiple non-crossing quantile fields, and as smooth probability density functions. They are modeled using Compositional Fixed Rank Kriging, Multiple Quantile Spatial Regression, and a Bayes Space Alignment with Fixed Rank Kriging, respectively. We apply the three approaches to daily PM\textsubscript{10} measurements collected across Northern Italy between 2018 and 2022 to estimate spatially continuous pollutant distributions and assess regulatory exceedance risk. \authors{Spatial block cross-validation shows that all three methods outperform Universal Block Kriging in predicting exceedance probabilities. Beyond this improvement in predictive accuracy, modeling the full distribution provides a coherent framework from which multiple distributional summaries can be derived without fitting separate spatial models.}}
% max: 200 parole


\maketitle

%%%%%%%%%%%%%%%%%%%%%%%%%%%%%%%%%%%%%%%%%%%%%%%%%%%%%%%%%%%%%%%%%%%%%%%%%%%
\section{Introduction}  
\label{section:intro}

Air quality assessment increasingly relies on distributional summaries, such as expected concentrations, quantiles, and regulatory exceedance probabilities, to characterize population exposure and support environmental decision making. These quantities are all derived from the same underlying probability distribution, yet they are typically modeled separately using scalar geostatistical approaches. This strategy may overlook important features of the pollutant distribution, requires fitting independent models for each quantity of interest, and may lead to mutually inconsistent estimates.

Rather than modeling individual summaries, we focus on the underlying probability distribution from which these quantities are derived. Specifically, the objective of this work is to model spatially varying probability distributions through three complementary distributional representations. Different representations retain different amounts of distributional information and naturally support different inferential goals. We consider representations based on two-part compositions, multiple non-crossing quantiles, and continuous probability density functions, modeled through Compositional Fixed Rank Kriging (CFRK), Multiple Quantile Spatial Regression (MQSR), and Bayes Space Alignment with Fixed Rank Kriging (BSA-FRK), respectively. Together, these approaches provide a coherent distributional framework for spatial air quality assessment.

Among air pollutants, we focus on \PMten, which  is one of the most relevant indicators of air quality because of its well-documented impact on human health \citep{mukherjee2017world}.  \PMten consists of particles with aerodynamic diameter smaller than 10 $\mu$m that can penetrate the respiratory tract and contribute to both acute and chronic cardiovascular and respiratory diseases \citep{PopeDockery2006}. According to the World Health Organization, ambient air pollution was responsible for approximately 4.2 million premature deaths worldwide in 2019 \citep{WHO2024}, while epidemiological studies have shown that the long-term burden of disease is primarily associated with sustained exposure to elevated pollutant concentrations \citep{dominici_health_2003,air_pollution_europe}. The spatial distribution of \PMten concentrations is driven by the interaction between anthropogenic emissions, including traffic and industrial activities, and local geomorphological and meteorological conditions \citep{pm10_sources_1}. This results in substantial spatial heterogeneity, making \PMten an ideal case study for investigating spatial distributional approaches. Northern Italy is particularly well suited for this purpose. The combination of intense urbanization, industrial activity, and the atmospheric conditions characterizing the Po Valley makes it one of the most polluted regions in Europe, with pronounced spatial variability in pollutant concentrations \citep{eea_home_2024}. Moreover, recent epidemiological evidence suggests that the cumulative distribution of pollutant exposure, rather than only its mean level, is strongly associated with adverse respiratory outcomes \citep{chen_cumulative_2022}. These characteristics make \PMten concentrations an ideal setting for investigating statistical methods that explicitly model the entire probability distribution, from which multiple distributional summaries can subsequently be derived.

Classical geostatistical and spatio-temporal models have been extensively used to analyze the spatial variability of \PMten\ concentrations \citep{angelici_2016,otto_2024}. These approaches typically model scalar distributional summaries, such as the expected concentration or other univariate characteristics of the pollutant distribution, and have proved effective for spatial interpolation and exposure assessment. Representative examples include Kriging for estimating intra-urban variation in \PMten\ exposure, generalized additive mixed models combining penalized splines with spatial random effects \citep{Paciorek2007}, and Bayesian spatio-temporal models based on the Stochastic Partial Differential Equation (SPDE) approach and Integrated Nested Laplace Approximation (INLA) \citep{Cameletti2013,rue_approximate_2009,lindgren2011explicit}.
Despite their success, these approaches generally model individual distributional summaries separately, rather than the underlying probability distribution from which they are derived. \authors{While convenient, separate scalar summaries may fail to capture the richness and heterogeneity of pollutant distributions across space, overlooking important distributional features such as asymmetry, tail behavior, and spatially varying variability. Moreover, fitting independent scalar models for different quantities of interest may lead to mutually inconsistent estimates.}

These considerations motivate an interest in statistical methodologies that treat the probability distribution itself as the primary object of analysis, rather than focusing on individual distributional summaries. By modeling the entire distribution, it becomes possible to derive expected concentrations, quantiles, exceedance probabilities, and other distributional summaries within a single mathematically coherent framework. At the same time, the complete probability distribution provides a richer characterization of pollutant exposure by accounting for features such as asymmetry, tail behavior, and spatially varying variability, which are directly relevant for environmental monitoring and risk assessment.

\authors{Recent years have witnessed a growing interest in the statistical analysis of complex data objects, including probability distributions. Several methodological frameworks have been proposed to represent and analyze distribution-valued data. One line of research transforms probability density functions into elements of Hilbert spaces through suitable mappings, such as log-hazard and log-quantile transformations, thereby enabling the application of functional data analysis techniques \citep{petersen2016functional}. Alternative approaches treat distributions directly as random objects in non-Euclidean spaces, including Bayes spaces and Wasserstein spaces, extending statistical methodology beyond the Euclidean setting \citep{egozcue2006hilbert, VanDenBoogart_Egozcue_Pawlowsky_Glahn_2011,gouet2015geodesic,HRON2016330,panaretos2019statistical}. Related developments have also emerged within the symbolic data analysis literature, where histogram- and distribution-valued observations are treated as complex statistical objects \citep{billard2006symbolic,noirhomme2011far,irpino2015basic}. In parallel, the functional geostatistics literature has extended classical kriging to function-valued spatial data \citep{Delicado2010,Giraldo2011,menafoglio2013universal,giraldo2014kriging,menafoglio2016kriging,Ignaccolo2018}. More recently, these ideas have been generalized to accommodate increasingly complex data objects, including probability density functions and other observations lying in non-Hilbert spaces \citep{menafoglio2016class,menafoglio2021kriging}. Comprehensive reviews of these developments are provided by \citet{menafoglio2017statistical}.}

Building on these methodological developments, we propose three spatial approaches for modeling probability distributions of \PMten concentrations. The proposed framework is applied to daily observations collected across Northern Italy between 2018 and 2022 to estimate spatially varying pollutant distributions and assess regulatory exceedance risk. \authors{A spatial block cross-validation study demonstrates that explicitly modeling the full probability distribution improves predictive accuracy even when interest is restricted to a single distributional summary. At the same time, the proposed approaches provide a mathematically coherent framework from which expected concentrations, quantiles, exceedance probabilities, and other distributional summaries can be derived without fitting separate spatial models.}

The remainder of the paper is organized as follows. Section~\ref{section:data} introduces the \PMten concentration dataset used in the case study. Section~\ref{sec:data-representations} presents the three distributional representations considered in this work. Section~\ref{section:models} introduces the corresponding statistical models for spatial prediction. Section~\ref{section:results} presents the results obtained for \PMten concentrations across Northern Italy. \authors{Section~\ref{subsection:CV} evaluates the predictive performance of the proposed approaches through a spatial block cross-validation study, comparing them with Universal Block Kriging for predicting exceedance probabilities.} Finally, Section~\ref{section:discussions} summarizes the main findings, discusses the practical implications of the proposed framework, and outlines directions for future research.

%%%%%%%%%%%%%%%%%%%%%%%%%%%%%%%%%%%%%%%%%%%%%%%%%%%%%%%%%%%%%%%%%%%%%%%%%%%
\section{Air quality data in Northern Italy} 
\label{section:data}

\begin{figure}
\centering
\caption{Spatial distribution of the monitoring stations measuring \PMten concentrations in Northern Italy during the period 2018--2022. The inset enlarges the area surrounding Trieste and reports the \PMten time series recorded at two monitoring stations. The time series in the top of this inset highlights an extreme \PMten episode associated with a wildfire on the Karst Plateau (Friuli-Venezia Giulia, July 2022), illustrating the type of transient event removed by the preprocessing procedure.}
\label{fig:monitoring-stations}
\vspace{7pt}
\includegraphics[width=0.9\linewidth]{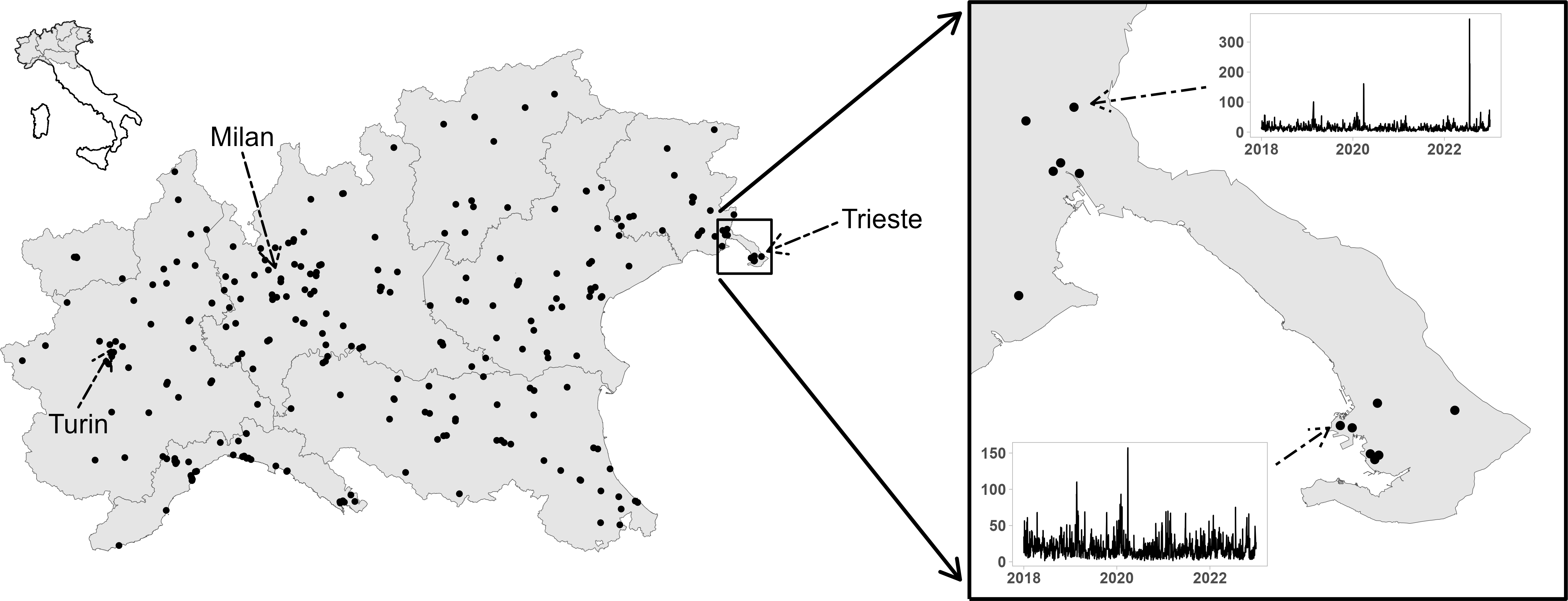}
\end{figure}

We analyze geo-referenced time series of daily average \PMten\ concentrations collected between 2018 and 2022 from 266 monitoring stations distributed across the study region. Figure~\ref{fig:monitoring-stations} shows the spatial distribution of the monitoring network together with the locations of the two largest cities, Milan and Turin. Since our objective is to characterize the structural air quality of the region rather than transient pollution episodes, we apply a spatially adaptive trimming procedure to the empirical \PMten\ distributions. Extremely high concentrations may reflect short-lived events, such as fireworks, bonfires, wildfires, agricultural activities, and dust storms \citep{pm10_sources_2}, whereas unusually low values may arise from measurement errors or sensor malfunctions. Such observations are not representative of the long-term air quality conditions that are the focus of the present study. Figure~\ref{fig:monitoring-stations} reports an example of a transient pollution episode. The trimming procedure is based on station-specific thresholds derived from the estimated 0.01 and 0.99 spatial quantile fields (\(Q_1\) and \(Q_{99}\)), computed using the spatial quantile regression approach proposed by \citet{castiglione_2025}. Observations falling outside the corresponding local quantile range are removed before constructing the empirical distributions. Appendix~\ref{section:appendix-a} provides full details of the preprocessing procedure.

\begin{figure}
\centering
\caption{
\authors{Spatial distribution of the covariates used in the analysis of \PMten concentrations. Left: population density at the municipal level, displayed using a square-root scale. Right: altitude, represented as a continuous raster field.}}
\label{fig:covariates}
\vspace{7pt}
\includegraphics[width=0.495\linewidth]{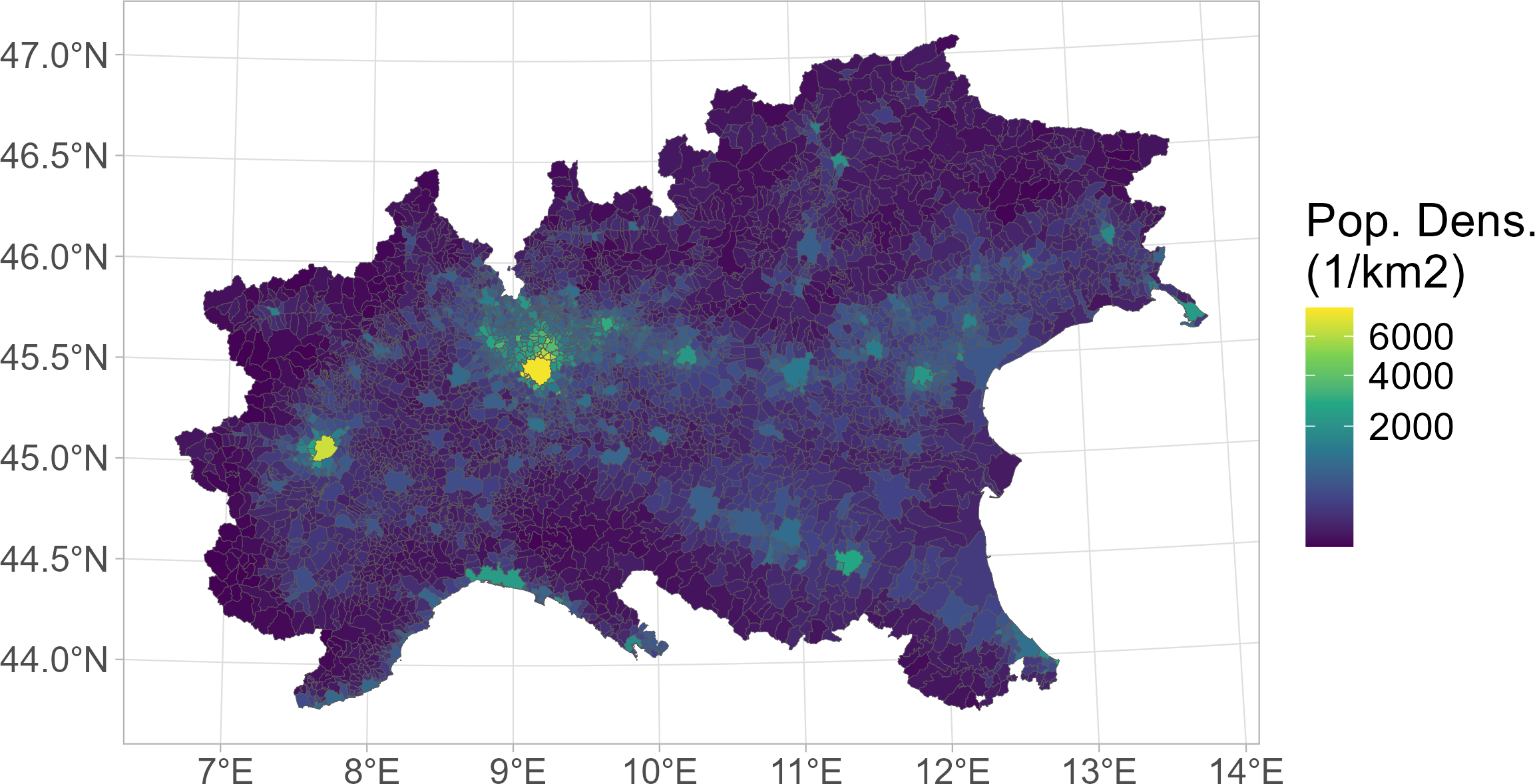}
%\hspace{0.03\linewidth}
\includegraphics[width=0.495\linewidth]{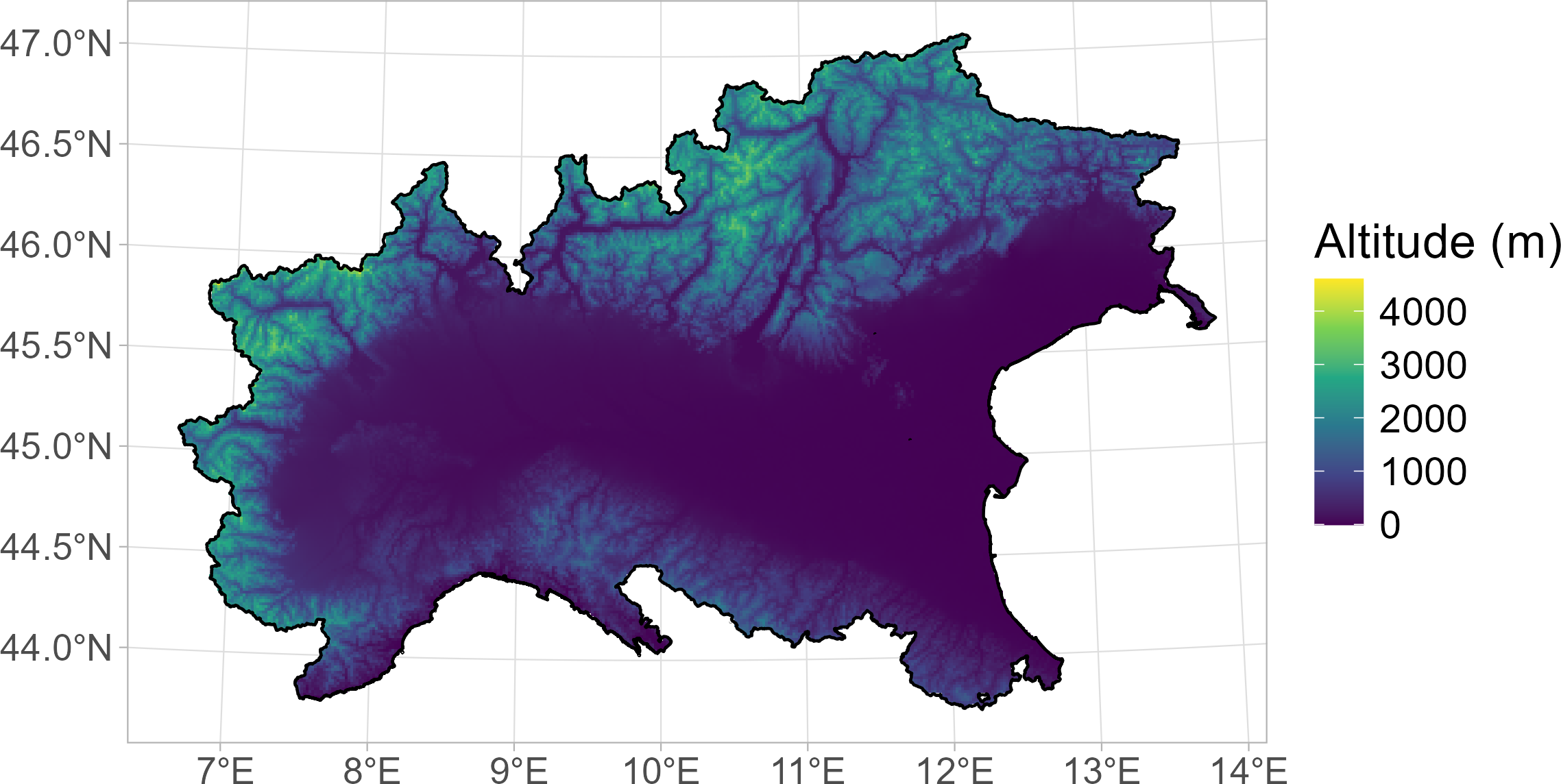}
\end{figure}

The proposed models include population density and altitude as spatial covariates, to account for the anthropogenic and geomorphological drivers of \PMten concentrations. Population density at the municipal level is obtained from the 2021 Census released by the Italian National Institute of Statistics (ISTAT; \url{https://www.istat.it/notizia/basi-territoriali-e-variabili-censuarie/}), whereas altitude is derived from the Copernicus Digital Elevation Model (DEM) distributed by Eurostat (\url{https://www.eea.europa.eu/en/datahub/datahubitem-view/d08852bc-7b5f-4835-a776-08362e2fbf4b}). Figure~\ref{fig:covariates} displays the spatial distribution of the two covariates. Population density highlights the major urban areas, including Milan, Turin, and their surrounding metropolitan regions. Because anthropogenic pressure varies smoothly over space, we apply the spatial downscaling approach of \citet{sangalli_spatial_2013,Azzimonti2015} to transform the municipality-level data into a continuous population density field. The elevation map shows the Alpine chain along the northern and western boundaries of the study region. This mountain barrier limits the inflow of air masses from Northern and Western Europe and favors atmospheric conditions, such as thermal inversions, that reduce air circulation over the Po Valley and promote pollutant accumulation.

%%%%%%%%%%%%%%%%%%%%%%%%%%%%%%%%%%%%%%%%%%%%%%%%%%%%%%%%%%%%%%%%%%%%%%%%%%%

\section{Representing Spatially Varying Distributions}
\label{sec:data-representations}

\begin{figure}
\caption{
\authors{Three representations of the local distribution of \PMten concentrations, providing progressively richer descriptions of the underlying distribution. Left: two-part composition based on the regulatory threshold of $50\,\mu g/m^3$, targeting the exceedance probability. Center: multiple-quantile representation, targeting a sequence of conditional quantiles that provide a piecewise approximation of the distribution. Right: continuous density representation, targeting the entire probability density function.}}
\label{fig:data-representations}
\centering
\includegraphics[width=.9\linewidth, trim={0.5cm, 0, 0, 0}, clip]{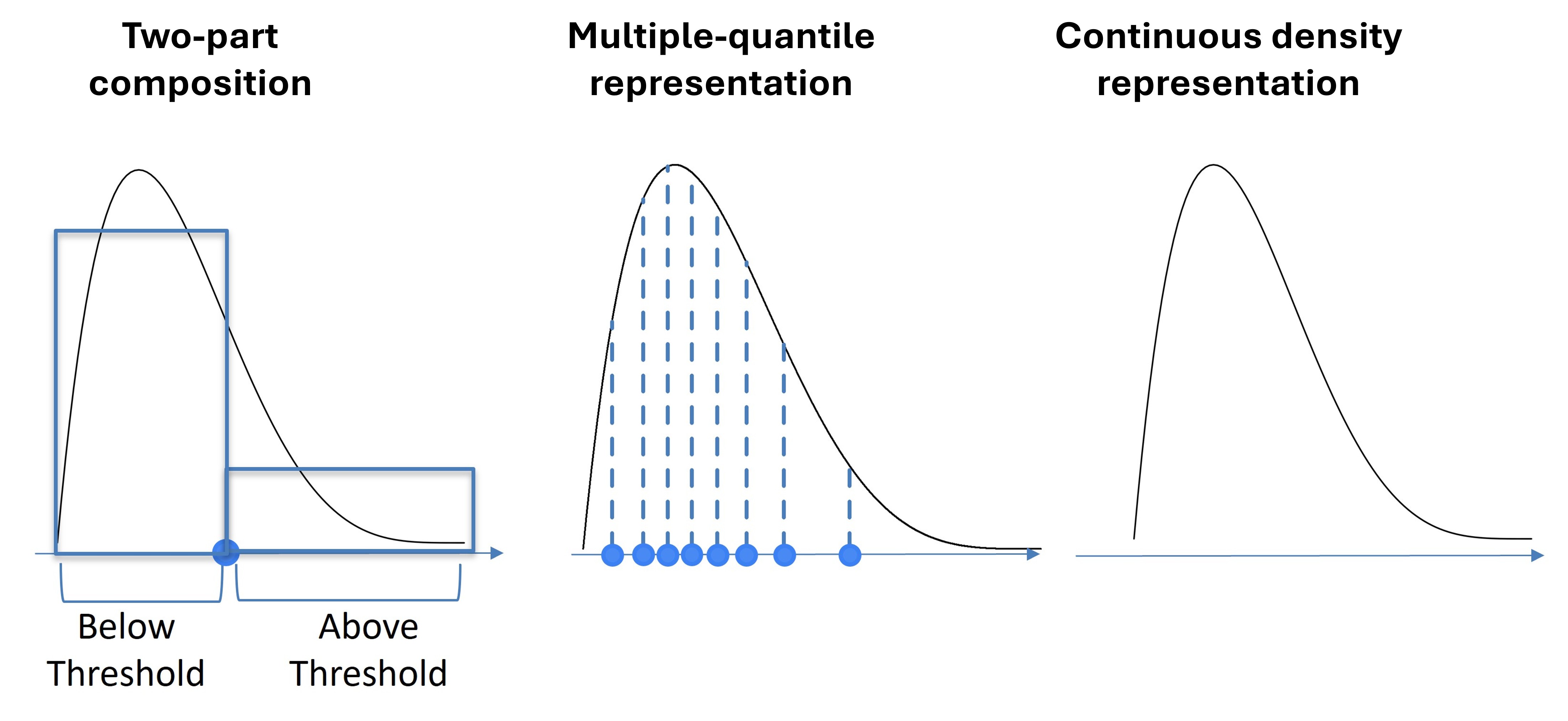}
\end{figure}

The object of interest in this work is the probability distribution of \PMten concentrations at each spatial location. Different representations of this distribution naturally retain different amounts of distributional information and support different inferential goals. We therefore consider three representations of progressively finer distributional resolution, illustrated in Figure~\ref{fig:data-representations}, which form the basis of the spatial models introduced in Section~\ref{section:models}.

\authors{
The first representation is \emph{compositional}. The support of the distribution is partitioned into two classes separated by the regulatory threshold of $50\ \mu g/m^3$, yielding a two-part composition that summarizes the probability of complying with, or exceeding, the regulatory limit. This threshold plays a central role in European air quality legislation, where the daily average concentration of \PMten should not exceed $50\ \mu g/m^3$ more than $35$ times per calendar year \citep{eea_annual_limit2024}. The resulting representation directly targets exceedance risk while providing a coarse approximation of the underlying distribution. The second representation is based on \emph{multiple quantiles}. Rather than partitioning the support through a fixed regulatory threshold, it characterizes the distribution by a sequence of increasing quantiles. Consecutive quantiles define probability classes, yielding a piecewise approximation of the underlying distribution that retains substantially richer distributional information while preserving a direct probabilistic interpretation. The third representation models the \emph{entire probability density function}. This provides the most informative description of the pollutant distribution, from which any distributional summary, including expected concentrations, quantiles, and exceedance probabilities, can be derived. Density estimation and spatial smoothing are carried out within a Functional Data Analysis framework.
}

These three representations provide progressively richer descriptions of the same underlying distribution. In the next section, we introduce the statistical models adopted to estimate and spatially predict each representation.

\section{Statistical models}
\label{section:models}

Consistent with the three representations of the pollutant distribution introduced in the previous section, we now present the corresponding statistical models for estimating and spatially predicting them.

We adopt the following notation throughout the section. Let $\mathcal{D}\subset\mathbb{R}^2$ denote the spatial domain of interest, which in our application corresponds to Northern Italy, and let $\{\mathbf{s}_i\}_{i=1}^n\subset\mathcal{D}$ be the set of spatial locations of the $n$ monitoring stations. At each location $\mathbf{s}_i$, we observe the vector of daily average \PMten concentrations $\mathbf{y}_i\in\mathbb{R}^{n_i}$, viewed as a realization of the random vector $\mathbf{Y}_i\in\mathbb{R}^{n_i}$, where $n_i$ denotes the length of the corresponding time series. We further denote by $\mathbf{x}_i\in\mathbb{R}^q$ the vector of $q\ge1$ covariates observed at location $\mathbf{s}_i$.

\subsection{Compositional Fixed Rank Kriging}
\label{section:CFRK}
% modello CFRK

Our first approach is Compositional Fixed Rank Kriging (CFRK), a hierarchical linear mixed-effects model for non-stationary and multiscale compositional data. CFRK extends the seminal Fixed Rank Kriging (FRK) methodology of \citet{Cressie2008} to compositional responses. While FRK was originally developed for large-scale spatial prediction of Euclidean responses and is now widely used in geostatistical applications \citep[see][]{ZammitMangion2021, Sainsbury-Dale2024}, CFRK adapts this framework to compositional data by exploiting the Aitchison geometry of the simplex.

Following the compositional representation introduced in the previous section, we represent the local distribution through the composition defined by the relative frequencies of observations falling within a fixed partition of the support of \PMten concentrations. Throughout this work, we consider the two-part partition induced by the regulatory threshold of \(50\ \mu g/m^3\), namely the intervals \(I_{[0,50)}\) and \(I_{[50,\infty)};\) this choice is motivated by the application, but, leveraging on multivariate extensions of FRK \citep{caringi2026frck}, the proposed methodology naturally extends to arbitrary finite partitions of the support, yielding compositional vectors with any finite number of parts. Accordingly, at each location $\mathbf{s}_i$, we observe the vector of covariates $\mathbf{x}_i$ and the composition
\begin{equation*}
C_{\mathbf{Y}_i|\mathbf{s}_i, \mathbf{x}_i} = \Big(
|\{\mathbf{y}_i\in I_{[0,50)}\}|/n_i, \ |\{\mathbf{y}_i\in I_{[50,\infty)}\}|/n_i
\Big).
\end{equation*}
The compositions \(\{C_{\mathbf{Y}_i|\mathbf{s}_i,\mathbf{x}_i}\}_{i=1}^n\) are elements of the simplex ${\cal A}^2\subset\mathbb{R}^2$, equipped with the Aitchison geometry \citep{aitchison2003statistical}. The transformation
\[
ilr:{\cal A}^2\rightarrow\mathbb{R}
\]
defined by
\begin{equation*}
C^{ilr} = \frac{1}{\sqrt{2}}\;\text{log}\,\frac{c_1}{c_2}
\end{equation*}
for all \(C=(c_1,c_2)\in{\cal A}^2\) is an isometric isomorphism that maps ${\cal A}^2$ into \(\mathbb{R}\), and is equivalent to the \emph{isometric log-ratio} transformation \citep[see, e.g.,][]{Egozcue2003}.

We model the \emph{ilr}-transformed responses \(\{C_{\mathbf{Y}_i|\mathbf{s}_i, \mathbf{x}_i}^{ilr}\}_{i=1}^n\) through the Fixed Rank Kriging framework as
\begin{equation}\label{eq:frk-terms}
C_{\mathbf{Y}_i|\mathbf{s}_i, \mathbf{x}_i}^{ilr} = \mathbf{x}_i^\top\boldsymbol\beta + \nu_i + \xi_i,\quad i=1,\dots,n,
\end{equation}
where $\nu_i$ and $\xi_i$ denote zero-mean random effects operating at two distinct spatial scales:
\begin{enumerate}
\item $\nu_i$ represents the ``small-scale'' variation, with \(\mathbb{E}(\nu_i)=0\).
\item $\xi_i$ represents the ``fine-scale'' variation, modeled as homoscedastic uncorrelated noise with \(\mathbb{E}(\xi_i)=0\) and \(\mathrm{Var}(\xi_i)=\sigma_\xi^2I\).
\end{enumerate}
The random effects $\nu_i$ are modeled as a linear combination of $r$ basis functions $\Phi_j(\mathbf{s}):\mathcal{D}\to\mathbb{R}$ with random coefficients $\eta_j\in\mathbb{R}$, for \(j=1,\dots,r\). Let \(\boldsymbol\Phi(\mathbf{s})=(\Phi_{1}(\mathbf{s}),\Phi_{2}(\mathbf{s}),\dots,\Phi_{r}(\mathbf{s}))^\top\) and \(\boldsymbol\eta=(\eta_1,\eta_2,\dots,\eta_r)^\top.\) Then \(\nu_i=\boldsymbol\Phi(\mathbf{s}_i)^\top\boldsymbol\eta\), and Equation~\eqref{eq:frk-terms} becomes
\begin{equation}\label{eq:frk-model}
C_{\mathbf{Y}_i|\mathbf{s}_i, \mathbf{x}_i}^{ilr} = \mathbf{x}_i^\top\boldsymbol\beta + \boldsymbol\Phi(\mathbf{s}_i)^\top\boldsymbol\eta + \xi_i,\quad i=1,\dots,n.
\end{equation}
The basis functions are radially symmetric and take the form \(\Phi_j((\mathbf{s}-\mathbf{s}_{c_j})/\sigma_l)\), with centers \(\mathbf{s}_{c_j}\) distributed across the spatial domain \(\mathcal{D}\) and organized into multiple resolution levels, each characterized by its own scale parameter \(\sigma_l\). Let \(\boldsymbol\Upphi\) denote the matrix of basis functions \(\boldsymbol\Phi(\mathbf{s})\) evaluated at the monitoring locations \(\{\mathbf{s}_i\}_{i=1}^n\). The variance-covariance matrix of \(\{C_{\mathbf{Y}_i|\mathbf{s}_i, \mathbf{x}_i}^{ilr}\}_{i=1}^n\) is therefore
\begin{equation}
\text{Var}(C_{\mathbf{Y}_i|\mathbf{s}_i, \mathbf{x}_i,\boldsymbol{\eta}}^{ilr}) = {\boldsymbol\Upphi}^\top K\boldsymbol\Upphi + \sigma_\xi^2I,
\end{equation}
where \(K=\mathrm{Var}(\boldsymbol\eta)\in\mathbb{R}^{r\times r}\), parameterized through a block-exponential structure, is referred to as the fixed rank matrix.

Under the Gaussian assumptions on \(\xi_i\) and \(\boldsymbol{\eta}\), the unknown parameters are estimated through an iterative Expectation--Maximization (EM) procedure, as detailed in \citet{ZammitMangion2021}. All analyses were carried out using the \texttt{FRK} package for \texttt{R} \citep{frk_package}.

\begin{figure}[t]
    \caption{Visual representation of the hierarchical structure of CFRK. The bottom layer shows the computational grid of Basic Areal Units (BAUs), together with the footprints of the georeferenced monitoring stations (black squares). The intermediate layer represents observations with heterogeneous spatial supports, whereas the top layer corresponds to the prediction supports, which in our application are the municipalities. This hierarchical structure enables estimation at the BAU level while accommodating observations and predictions over different spatial supports. 
    }
    \label{fig:CFRK}
    \centering
    \includegraphics[width=0.9\linewidth]{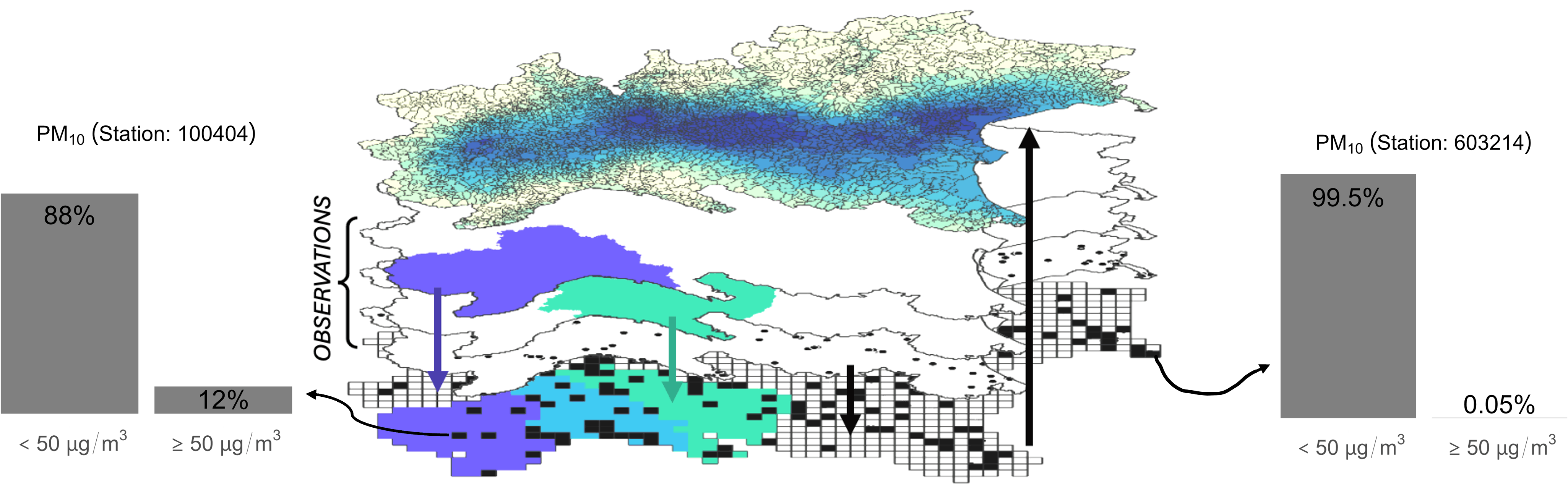}
\end{figure}
Fixed Rank Kriging (FRK) is implemented within a hierarchical framework consisting of a three-layer structure organized as follows:
\begin{enumerate}
    \item \textbf{Hidden layer:} a computational grid of Basic Area Units (BAUs), where the FRK model is implemented. For the analyses presented in this work, the BAUs grid covering the spatial domain \(\mathcal D\) has a cell size of 1.6 km, yielding a grid composed of $46010$ cells. A random selection of 502 BAU centers \(\{\mathbf{s}_{c_j}\}\) is then used to define 502 basis functions \(\Phi_j((\mathbf{s} - \mathbf{s}_{c_j}) / \sigma_l)\), organized across three resolution levels.
   \item \textbf{Observation layer:} data with potentially heterogeneous geometric support, where each dataset has its own footprint onto the BAUs. In this study, this layer corresponds to the BAUs grid, since each observation is generated by a monitoring station identified with a location \(\mathbf{s}_i\).
    \item \textbf{Prediction layer:} a third layer which, in the present study, corresponds to the lattice generated by the municipal boundaries. As in the observation layer, the target variable \(C_{\mathbf{Y}|\mathbf{s}, \mathbf{x}}^{ilr}\), evaluated at the BAU level, is integrated over each polygonal support by applying appropriate weights.
\end{enumerate}
Figure~\ref{fig:CFRK} illustrates the projection of the observation footprints onto the BAUs grid within the CFRK framework.
Finally, the estimate of the probability \(P(Y(\mathbf{s})\in I_{[50,\infty)})\) over the domain \(\mathcal{D}\) is obtained by inverting the prediction of the \emph{ilr}-transformed variable and integrating it over the municipalities.

\subsection{Multiple Quantile Spatial Regression}
\label{section:MQSRPDE}
% modello MQSRPDE
Following the multiple-quantile representation introduced in Section~\ref{sec:data-representations}, the second approach is based on Multiple Quantile Spatial Regression (MQSR), proposed by \citet{de_sanctis_2025} and implemented in the \texttt{R} package \texttt{fdaPDE} \citep{fdaPDE}. MQSR characterizes spatially varying distributions through a sequence of probability classes defined by increasing quantile levels.
For a grid of quantile levels \(0 < \alpha_1 < \dots < \alpha_r < 1\), we estimate the corresponding quantile fields over the spatial domain. Specifically, the quantiles of the responses \(\{\mathbf{Y}_i\}_{i=1}^n\), conditional on the spatial locations \(\{\mathbf{s}_i\}_{i=1}^n\) and covariates \(\{\mathbf{x}_i\}_{i=1}^n\), are jointly modeled as
\begin{equation*}
Q_{\mathbf{Y}_i|\mathbf{s}_i,\mathbf{x}_i}(\alpha_j) = \mathbf{x}_i^\top \boldsymbol{\beta}_j + f_j(\mathbf{s}_i), \quad i=1,\dots,n, \ \ j=1,\dots,r,
\end{equation*}
where \(\boldsymbol{\beta}_j \in \mathbb{R}^q\) represents the global linear effect of the covariates at quantile level \(\alpha_j\), and \(f_j:\mathcal{D}\rightarrow\mathbb{R}\) is a deterministic spatial field describing the spatial variability of the response at the same quantile level. The unknown regression coefficients \(\boldsymbol{\beta}=(\boldsymbol{\beta}_1,\dots,\boldsymbol{\beta}_r)\) and spatial fields \(\boldsymbol{f}=(f_1,\dots,f_r)\) are estimated simultaneously across the \(r\) quantile levels by solving a penalized optimization problem. Specifically, we minimize the estimation functional
\begin{equation}
   J_P(\boldsymbol{\beta}, \boldsymbol{f}) = \mathcal{L}(\boldsymbol{\beta}, \boldsymbol{f}) + P(\boldsymbol{f}) + \mathcal{C}(\boldsymbol{\beta}, \boldsymbol{f}).
\label{eq:objective}
\end{equation}

The first term, \(\mathcal{L}(\boldsymbol{\beta}, \boldsymbol{f})\), is the quantile regression loss based on the pinball function, which measures the empirical fitting error for the \(\alpha_j\)-quantile at each observation \citep[see, e.g.,][]{bassett_koenker_1982},
\begin{equation*}
   \mathcal{L}(\boldsymbol{\beta}, \boldsymbol{f}) = \frac{1}{n} \sum_{j=1}^{r} \sum_{i=1}^{n} \sum_{k=1}^{n_i} \rho_{\alpha_j}(y_k - \mathbf{x}_i^\top \boldsymbol{\beta}_j - f_{j}(\mathbf{s}_i)).
\end{equation*}

The second term, \(P(\boldsymbol{f})\), is a roughness penalty that penalizes deviations of each spatial field \(f_j\) from a problem-specific Partial Differential Equation (PDE), thereby incorporating prior knowledge about both the spatial structure of the smooth component and the geometry of the spatial domain. For simplicity, we consider an isotropic diffusion equation with zero forcing term \citep[see][]{sangalli_spatial_2013, sangalli2021spatial, castiglione_2025}:
\begin{equation}
\label{eq:P(f)}
   P(\boldsymbol{f}) = \sum_{j=1}^r \lambda_j \int_{\mathcal{D}}( \Delta f(\mathbf{s}) )^2 \,d\mathbf{s}.
\end{equation}
More generally, the differential regularization can be defined through a spatially varying PDE describing the underlying physics of the phenomenon under study, such as air flows for pollutant dispersion or sea currents for oceanographic applications, as illustrated by \citet{castiglione_2025} and \citet{de_sanctis_2025}. 

Finally, the third term, \(\mathcal{C}(\boldsymbol{\beta}, \boldsymbol{f})\), is a crossing penalty that penalizes pairs of quantiles \(\{Q(\alpha_j),Q(\alpha_{j+1})\}\) violating the monotonicity constraint \(Q(\alpha_j)\le Q(\alpha_{j+1})\). This term is defined as
\begin{equation}
\label{eq:Crossing}
   \mathcal{C}(\boldsymbol{\beta}, \boldsymbol{f}) = \gamma \sum_{j=1}^{r-1} \sum_{i=1}^{n} \max\{0, \varepsilon-\mathbf{x}_i^\top (\boldsymbol{\beta}_{j+1}-\boldsymbol{\beta}_j)-(f_{j+1}(\mathbf{s}_i) - f_{j}(\mathbf{s}_i))\}.
\end{equation}

The tuning parameters \(\{\lambda_j\}_{j=1}^{r}\subset\mathbb{R}^+\) in Equation~\eqref{eq:P(f)}, together with \(\gamma>0\) and \(\varepsilon>0\), can be selected using the data-driven procedure described in \citet{de_sanctis_2025}.

\begin{figure}[tb]
    \caption{Finite element discretization underlying the MQSR model. Left: triangulation of the Northern Italy spatial domain. Right: example of a linear finite element basis function defined on the triangulation. The compact support of the basis function induces sparsity in the resulting discretization matrices.}
    \label{fig:FEM}
    \centering\includegraphics[width=0.9\linewidth]{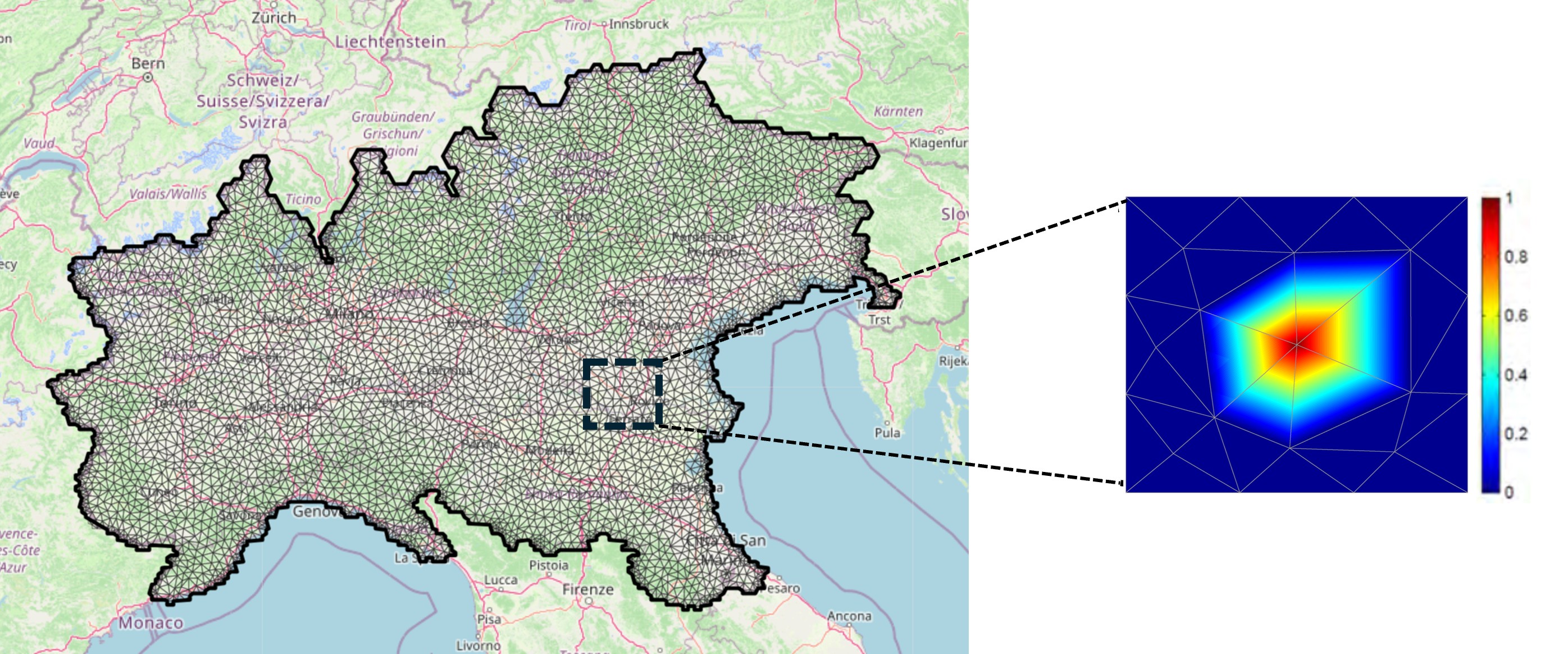}
\end{figure}

\authors{Unlike CFRK, MQSR does not require the specification of a compositional metric, since the estimation problem is formulated as the minimization of Equation~\eqref{eq:objective} within a fully nonparametric framework.} As for the CFRK model, the MQSR approach also requires an appropriate discretization scheme to estimate the spatial quantile fields \(\{f_j\}_{j=1}^{r}\). Specifically, the minimization of Equation~\eqref{eq:objective} is carried out using the Finite Element Method on a triangulation of the spatial domain \citep[see, e.g.,][]{brezzi_mixed_1991}. The left panel of Figure~\ref{fig:FEM} shows a triangulation of Northern Italy with \(24407\) vertices. This discretization naturally accommodates spatial domains with complex geometries, an important feature whenever the shape of the domain influences the phenomenon under study, as, for example, when modeling quantities measured over water bodies. The right panel displays a piecewise linear finite element basis function defined on the triangulation, illustrating its compact support, which induces sparse discretization matrices for the resulting estimation problem \citep{de_sanctis_2025}. This sparsity is essential from a computational perspective, since the estimation procedure requires the iterative solution of large linear systems. 

Specifically, the proposed estimation strategy relies on a Maximization--Minimization algorithm \citep[see, e.g.,][]{hunter_quantile_mm_2000}, for which, at each iteration, \authors{the updates are given by:}
\authors{
\begin{equation*}
\begin{split}
    &\hat{\boldsymbol{\beta}} = \left(X^\top W X\right)^{-1} X^\top W (\mathbf{y} - \Psi \hat{\mathbf{f}}), \\ 
    &\hat{\mathbf{f}} = \frac{1}{n} \left(\frac{1}{n} \Psi^\top Q \Psi + P \right)^{-1} \Psi^\top Q \mathbf{y}, 
\end{split}
\end{equation*}
where \(X\) and \(\Psi\) are the design matrices whose \(i\)-th rows contain the covariate vector \(\mathbf{x}_i\) and the evaluations of the finite element basis functions at location \(\mathbf{s}_i\), respectively; \(Q=W(I-X(X^\top W X)^{-1}X^\top W)\);  \(P\) is the discrete finite element counterpart of the smoothing penalty in Equation~\eqref{eq:P(f)}; \(W\) and \(\mathbf{z}\) denote the weight matrix and the vector of pseudo-observations, respectively, as defined by the Maximization--Minimization algorithm.
}

\authors{The monotonicity of the estimated quantiles enables the reconstruction of the probability density function at any spatial location. As discussed in Section~\ref{section:intro}, this provides a key methodological advantage over classical approaches: a single model yields coherent estimates of all distributional summaries, avoiding the need to fit separate spatial models for individual quantities of interest.}

\subsection{Bayes Space Alignment for Fixed Rank Kriging}
\label{section:BSAFRK}

Following the continuous density representation introduced in Section~\ref{sec:data-representations}, the third approach, named Bayes Space Alignment for Fixed Rank Kriging (BSA-FRK), treats probability density functions as functional data. BSA-FRK integrates Functional Data Analysis (FDA) techniques with the FRK and MQSR methodologies introduced in the previous sections to estimate and spatially predict probability density functions.

The classical Functional Data Analysis (FDA) framework embeds functional observations in \(L^2(I)\), the space of square-integrable real-valued functions with support \(I\); see, for example, \citet{Ignaccolo2018}. Probability density functions, however, constitute a special class of functional data, and representing them in \(L^2(I)\) ignores their defining constraints, namely non-negativity and unit integral. This limitation is well recognized in the literature, and several authors \citep[see][]{menafoglio2014kriging, menafoglio2016class, HRON2016330, scimone2022look, martinez2025modelling, vskorvna2026compositional} have proposed methodologies specifically designed for the analysis of spatially varying densities. Following this line of research, we embed the \PMten concentration densities in the Bayes space \(B^2(I)\) \citep{egozcue2006hilbert, VanDenBoogart_Egozcue_Pawlowsky_Glahn_2011}, which naturally accommodates these constraints and extends the Aitchison geometry mentioned in Section~\ref{section:CFRK} 
to the functional setting.

Let \(B^2(I)\) denote the space of positive real-valued functions with support \(I\subset\mathbb{R}\), unit integral, and satisfying \(\int_I (\log f(t))^2\,\mathrm{d}t<\infty\). The Bayes space \(B^2(I)\) forms a vector space when equipped with the following operations,
\[
(f \oplus g)(t) = \frac{f(t)g(t)}{\int_I f(s)g(s)\, \text{d}s}, \qquad (\alpha \odot f)(t) = \frac{f^\alpha(t)}{\int_I f^{\alpha}(s)\, \text{d}s},
\]
for \(f,g\in B^2(I)\) and \(\alpha\in\mathbb{R}\). An inner product between pairs of elements of \(B^2(I)\) is defined by
\[
\langle f, g \rangle_{B} = \frac{1}{2|I|}\iint_I \log \frac{f(t)}{f(s)}\log \frac{g(t)}{g(s)}\, \text{d}t\text{d}s.
\]
where \(|I|\) denotes the length of the interval \(I\). \citet{egozcue2006hilbert} showed that the vector space \(B^2(I)\), equipped with the inner product \(\langle f, g \rangle_{B}\), is a separable Hilbert space and, therefore, isomorphic to \(L^2(I)\). An isometric isomorphism between \(B^2(I)\) and \(L_0^2(I),\) the closed subspace of \(L^2(I)\) consisting of square integrable functions with zero integral, is provided by the \emph{centered log-ratio (clr)} transformation \citep{van2014bayes}:
\begin{equation}
clr(f)(t) = \log f(t) - \frac{1}{|I|}\int_I \log f(s)\, \text{d}s.
\label{eq:clr}
\end{equation}
The \emph{clr} transformation provides a bridge between \(B^2(I)\) and \(L^2(I)\), enabling the use of standard FDA techniques while preserving the geometric structure of probability density functions.

Despite its theoretical advantages, the Bayes space \(B^2(I)\) has one important limitation: all density functions must share the same support \(I\). This assumption is unrealistic for the present application because the support of the local \PMten distributions varies substantially across the study region. For example, monitoring stations in the mountainous regions of Aosta Valley and Trentino--Alto Adige typically record \PMten concentrations below \(80\,\mu g/m^3\), whereas stations in the agricultural and heavily industrialized Po Valley frequently exceed \(120\,\mu g/m^3\).

We address this limitation through an alignment strategy that maps all densities onto a common support. A probability density function can be characterized by two complementary features: its shape and its support. The support identifies the domain where the density is strictly positive, whereas the shape captures distributional characteristics that are invariant under location and scale transformations, such as skewness, modality, and the relative concentration of probability mass. Assuming that shape and support can be modeled separately, the common-support limitation of the Bayes space can be overcome by aligning all densities onto the interval \([0,1]\).
The trimming procedure introduced in Section~\ref{section:data} implies that, in practice, the relevant \PMten concentrations recorded at the \(i\)-th monitoring station, \(\lbrace y_{ij} \rbrace_{j=1}^{n_i}\), always lie between \(Q_1(i)\) and \(Q_{99}(i)\). The aligned observations, denoted by \(\tilde{y}_{ij}\), are then defined as
\begin{equation}
\tilde{y}_{ij} = \frac{y_{ij} - Q_1(i)}{Q_{99}(i) - Q_1(i)}, \qquad j = 1, \dots, n_i.
\label{eq:align-values}
\end{equation}
By construction, \(\tilde{y}_{ij}\in[0,1]\) for every observation and every monitoring station. Consequently, for \(i=1,\dots,n\), the aligned \PMten concentrations \(\{\tilde{y}_{ij}\}_{j=1}^{n_i}\) can be regarded as a sample from a distribution with density \(\tilde f_i\) supported on \([0,1]\).
The densities \(\tilde{f}_i\) are estimated using smoothing splines in \(B^2([0,1])\), as described by \citet{machalova2016preprocessing} and implemented in the \texttt{R} package \texttt{robCompositions} \citep{robCompositions}. The regularization parameters controlling the smoothness of the estimated densities are selected by cross-validation.

To spatially interpolate the aligned densities \(\tilde f_i\), we first map them into \(L^2([0,1])\) through the \emph{clr} transformation. Following the Simplicial Principal Component Analysis framework of \citet{HRON2016330}, the \emph{clr}-transformed densities are then represented in a finite-dimensional space by means of Functional Principal Component Analysis in \(L^2([0,1])\). In our application, the first four principal components explain nearly 90\% of the total variability. The corresponding scores are subsequently spatially interpolated over the municipalities of Northern Italy using the standard FRK methodology of \citet{Cressie2008}, with population density and elevation as covariates. \authors{More precisely, the FRK model for the scores associated with the \(l\)-th principal component is given by}
\[
\authors{\gamma_l(\bm{s}_i) = \mathbf{x}_i^{T}\bm{\beta}_l + \nu_{i, l} + \xi_{i, l}, \quad i=1, \dots, n,}
\]
\authors{where \(\nu_{i,l}\) and \(\xi_{i,l}\) denote the small-scale and fine-scale variation, respectively, for the \(i\)-th monitoring station and the \(l\)-th principal component. Following the FRK framework described above, the functional PCA scores are predicted at the municipal level by integrating the values of \(\gamma_l(\mathbf{s})\) evaluated over the BAUs.}
The interpolated scores are then used to reconstruct the \emph{clr}-transformed density function for each municipality. The reconstructed density is subsequently mapped back to \(B^2([0,1])\) through the inverse \emph{clr} transformation, yielding the predicted aligned density \(\tilde f\) supported on \([0,1]\).
Finally, the predicted \PMten concentration densities on their original scale are obtained by inverting the alignment transformation in Equation~\eqref{eq:align-values} using the municipal-level estimates of the spatial quantile fields \(Q_1\) and \(Q_{99}\), obtained from MQSR over the entire Northern Italy domain. The resulting density functions provide spatial predictions of the full \PMten concentration distribution across the study region.

\section{Results}
\label{section:results}

This section presents the results obtained by applying the three methods introduced in Section~\ref{section:models} to the \PMten data described in Section~\ref{section:data}. Since the methods rely on different representations of distributional data, a direct comparison of their outputs is not straightforward. We therefore focus on a set of distributional summaries derived from the predicted objects that are directly relevant for air quality assessment.
\authors{A key advantage of the proposed approaches is that all these summaries can be derived from a single fitted model, without re-estimating model parameters. This overcomes a limitation of classical geostatistical approaches, which typically require fitting separate models for each distributional summary.}

The first distributional summary we consider is the exceedance probability map, denoted by \(p\), corresponding to the probability that the \PMten concentration exceeds the regulatory threshold of \(50\,\mu g/m^3\). To facilitate the comparison among the three approaches, all estimates are reported at the municipal level. CFRK predicts the exceedance probability map \(p\) on the BAU grid; municipal-level estimates are then obtained by aggregating BAU predictions using area-weighted averages based on the overlap between BAUs and municipal boundaries. MQSR estimates spatially continuous quantile fields, from which the exceedance probability map \(p\) is derived. The resulting exceedance probability map is then integrated over each municipality to obtain municipal-level estimates. Finally, BSA-FRK directly predicts \PMten concentration density functions at the municipal level. Exceedance probabilities are subsequently obtained by numerically integrating the predicted densities above the regulatory threshold.

\begin{figure}
    \centering
    \caption{First row: choropleth maps of the predicted probability that the daily \PMten concentration exceeds the regulatory threshold of \(50\,\mu g/m^3\) at the municipal level, estimated using the three methods described in Section \ref{section:models}\authors{: Compositional Fixed Rank Kriging (left), Multiple Quantile Spatial Regression (center), and Bayes Space Alignment with Fixed Rank Kriging (right).} \authors{Second row: choropleth maps of the block bootstrap standard deviation of the estimated exceedance probability for the three proposed approaches. Third row: choropleth maps of the corresponding block bootstrap bag width.} \\}
    \label{fig:pred-prob}
    
    \begin{tikzpicture}
        \node[anchor=south west, inner sep=0] (image) at (0,0) {\includegraphics[width=0.92\linewidth]{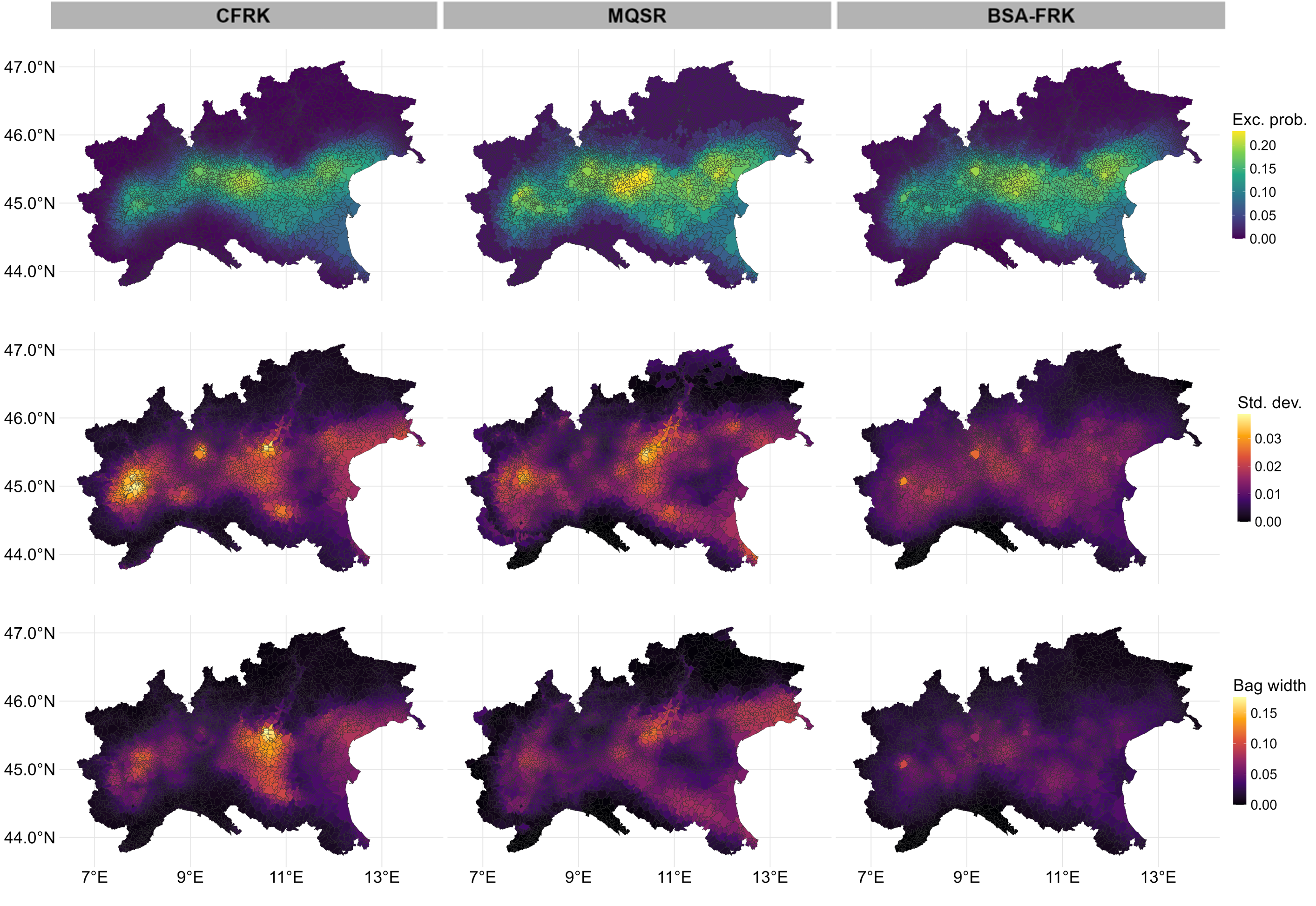}};
        \begin{scope}[x={(image.south east)},y={(image.north west)}]
            \node[rotate=90, anchor=center, font=\footnotesize\sffamily] at (-0.04, 0.81) {Exceedance probability};
            \node[rotate=90, anchor=center, font=\footnotesize\sffamily] at (-0.04, 0.50) {Standard deviation};
            \node[rotate=90, anchor=center, font=\footnotesize\sffamily] at (-0.04, 0.18) {Bag width};
        \end{scope}
    \end{tikzpicture}
\end{figure}

The top panels of Figure~\ref{fig:pred-prob} display the estimated exceedance probability \(p\) obtained with the three proposed methods. Despite their different distributional representations, all approaches reveal remarkably consistent spatial patterns across Northern Italy, identifying the Po Valley as the area at greatest risk of regulatory exceedance. This region is one of the main industrial and agricultural areas in Europe and is characterized by unfavorable orographic conditions that limit air circulation and promote the accumulation of \PMten. In contrast, the estimated exceedance probability decreases markedly in the Alpine region, along the Apennine chain, and near the Ligurian coast, where more favorable atmospheric conditions enhance pollutant dispersion.
\authors{The second and third rows of Figure~\ref{fig:pred-prob} summarize the uncertainty associated with the estimated exceedance probability \(p\). To quantify uncertainty, we adopt a block bootstrap procedure, a well-established approach for spatially and temporally correlated data \citep[see, e.g.,][]{franco2017bootstrap, zhu2004comparison}. Specifically, we generate \(50\) bootstrap replicates of the \PMten data using the \texttt{BlockBootID} function of the \texttt{R} package \texttt{ecostats} \citep{warton2022eco}, which implements a moving block bootstrap algorithm.
The second row of Figure \ref{fig:pred-prob} displays the standard deviation of the bootstrap estimates of the exceedance probability \(p\). We further summarize the bootstrap replicates through functional boxplots \citep[see, e.g.,][]{sun2011functional}, based on the Modified Band Depth (MBD) implemented in the \texttt{R} package \texttt{roahd} \citep{ieva2019roahd}. The third row of Figure \ref{fig:pred-prob} reports the corresponding bag width, defined as the width of the envelope containing the \(50\%\) deepest curves.
Visual inspection of the second and third rows suggests that BSA-FRK is generally associated with lower uncertainty than CFRK and MQSR. The latter two methods exhibit more localized uncertainty patterns, with larger variability over the central Po Valley and around the metropolitan area of Turin.}

\begin{figure}
    \centering
   \caption{Municipalities highlighted in red are those expected to exceed the regulatory \PMten threshold on more than 35 days per year\authors{, according to the three methods described in Section~\ref{section:models}: Compositional Fixed Rank Kriging (left), Multiple Quantile Spatial Regression (center), and Bayes Space Alignment with Fixed Rank Kriging (right).} Black markers indicate the monitoring stations.}
    \label{fig:anomalies}
    \includegraphics[width=\linewidth]{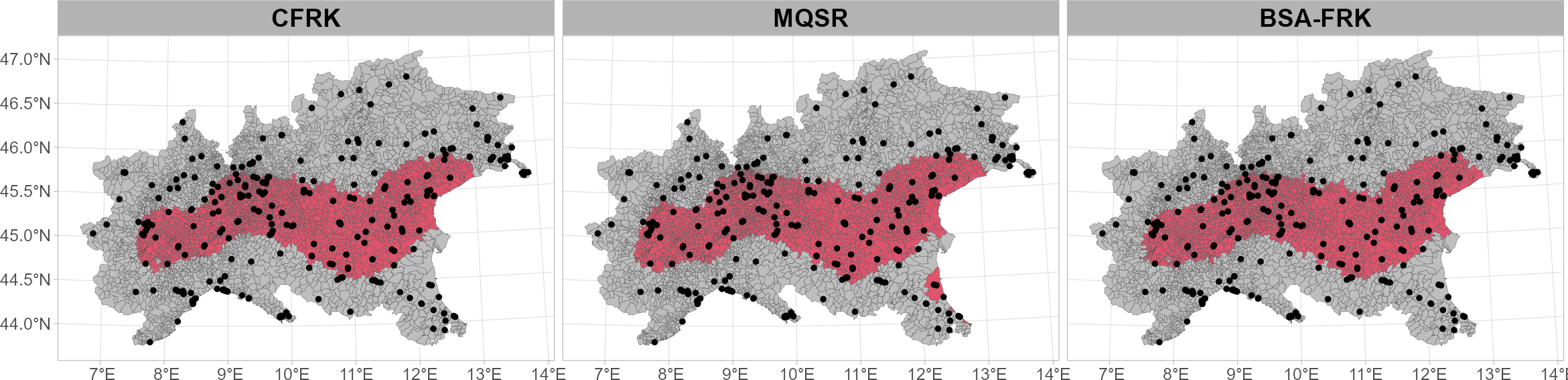}
\end{figure}

We next consider another indicator monitored by the EEA: the expected annual number of days with \PMten concentrations exceeding the regulatory threshold of \(50\,\mu g/m^3\). Figure~\ref{fig:anomalies} compares the three approaches in identifying municipalities where this threshold is exceeded for more than \(35\) days per year. Minor discrepancies appear only along the boundaries of the high-risk region. In particular, MQSR identifies a few additional municipalities in the south-eastern part of the Po Valley, including Ravenna, whereas CFRK and BSA-FRK produce nearly indistinguishable results. Despite these local differences, all three approaches reveal highly consistent spatial patterns, identifying the central Po Valley as the area with the highest expected frequency of regulatory exceedances.

\begin{figure}
    \centering
   \caption{Agreement among the three statistical approaches in identifying municipalities expected to exceed the regulatory \PMten threshold on more than 35 days per year. \authors{Black triangles mark the municipalities of Imperia, Ravenna, and Montichiari, each representing a different level of agreement among the three approaches.}
   }
    \label{fig:traffic-stop}
    \includegraphics[width=0.7\linewidth]{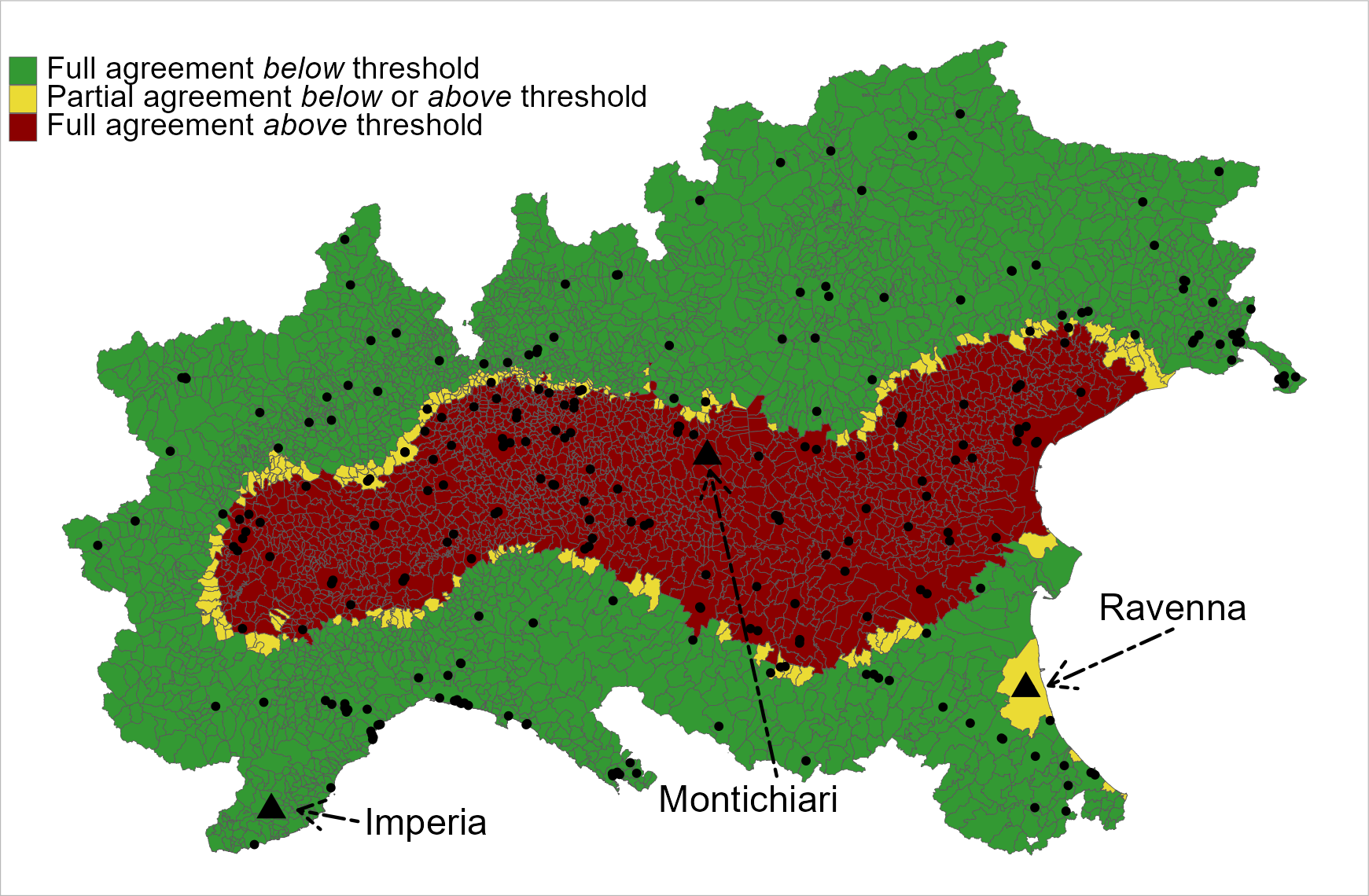}
\end{figure}

Figure~\ref{fig:traffic-stop} provides an alternative visualization of the agreement among the three approaches. Municipalities shown in green correspond to areas where all methods predict fewer than \(35\) exceedance days per year, whereas dark red identifies municipalities where all three methods agree in predicting regulatory exceedances. Yellow indicates municipalities for which the three approaches do not reach full agreement.
This representation further highlights the strong consistency among the methods. Agreement is nearly complete in both the core high-risk areas of the Po Valley (dark red) and the rural and mountainous regions (green), with disagreement confined to a limited number of municipalities along the boundary between these two areas (yellow).

\begin{figure}
    \centering
    \caption{Municipal-level estimates of the 95th percentile (\(q_{95}\)) of the daily \PMten concentration obtained using Multiple Quantile Spatial Regression (left) and Bayes Space Alignment with Fixed Rank Kriging (right). Compositional Fixed Rank Kriging is not included because its two-part compositional representation is defined by the fixed regulatory threshold and is therefore not suitable for estimating generic distributional summaries such as \(q_{95}\).}
    \label{fig:q95}
    \includegraphics[width=0.9\linewidth]{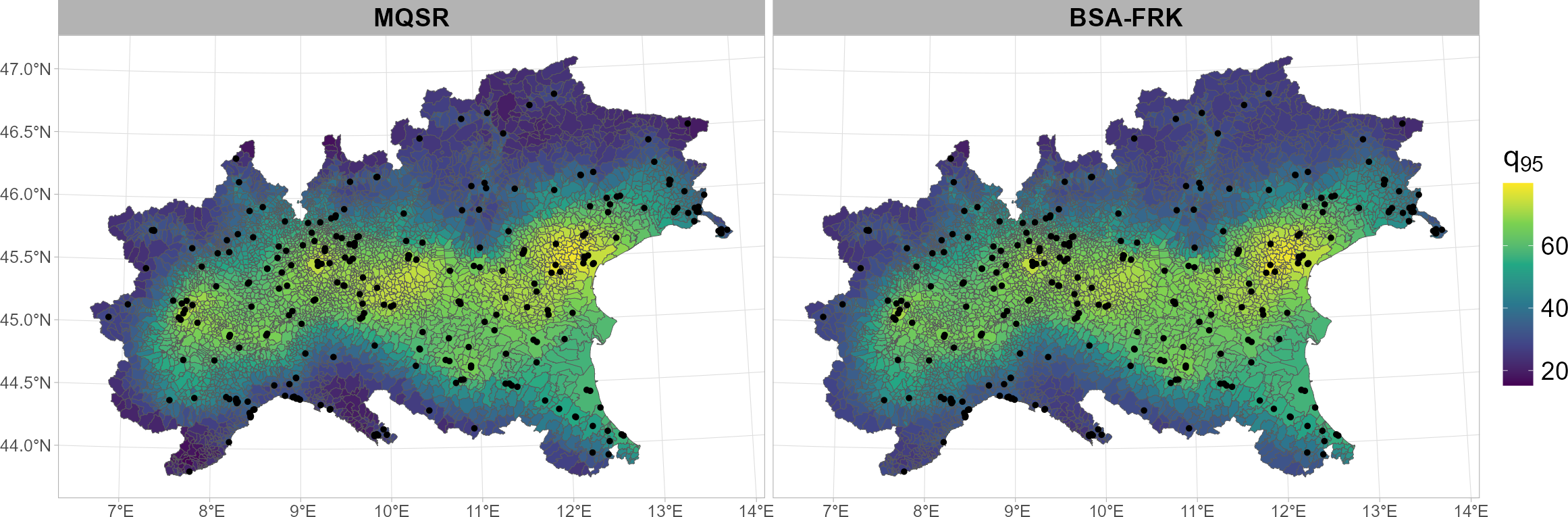}
\end{figure}

% q95 
Figure~\ref{fig:q95} displays the third distributional summary considered in this work, namely the 95th percentile (\(q_{95}\)) of the \PMten concentration, comparing the MQSR and BSA-FRK approaches. CFRK is not included in this comparison because its two-part compositional representation is defined by the fixed regulatory threshold of \(50\,\mu g/m^3\), making it unsuitable for estimating generic quantiles such as \(q_{95}\). Both MQSR and BSA-FRK reveal highly consistent spatial patterns for the 95th percentile, with the highest values concentrated over the Po Valley and progressively decreasing toward the Alpine region and the southern part of the study area.

\begin{figure}
    \centering
    \caption{Predicted \PMten concentration density functions obtained using Multiple Quantile Spatial Regression (left) and Bayes Space Alignment with Fixed Rank Kriging (right) for the municipalities of Imperia, Ravenna, and Montichiari. Colors correspond to the agreement regions shown in Figure~\ref{fig:traffic-stop}.}
    \label{fig:densities}
    \includegraphics[width=\linewidth]{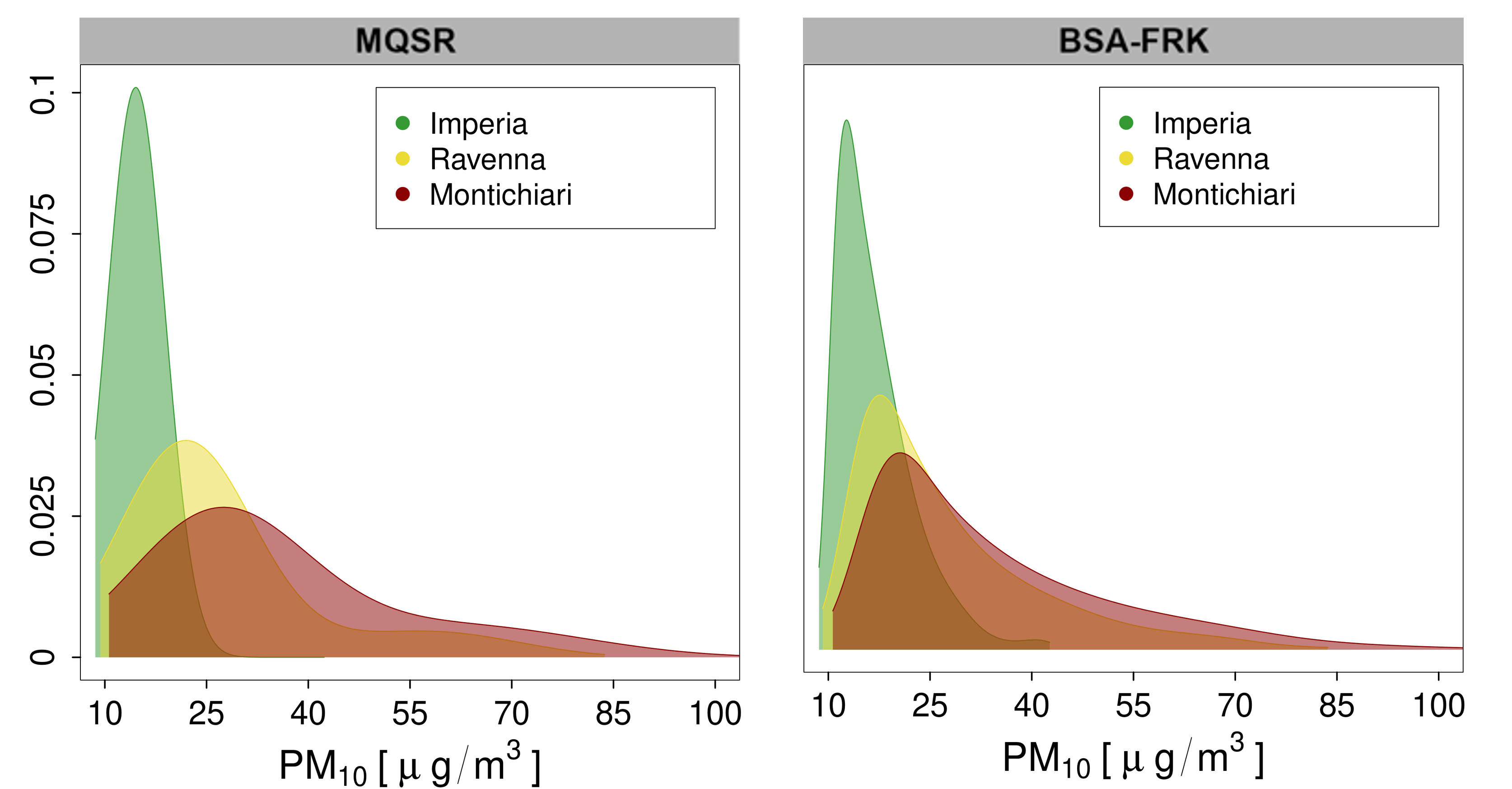}
\end{figure}

% densities 
Finally, Figure~\ref{fig:densities} displays the predicted \PMten concentration density functions for three municipalities representative of the different agreement scenarios identified in Figure~\ref{fig:traffic-stop}: Imperia, where all methods agree that the regulatory threshold is exceeded on fewer than 35 days per year (green region); Montichiari, where all methods predict more than 35 exceedance days per year (dark red region); and Ravenna, where the three approaches do not reach full agreement (yellow region).
Unlike the previous figures, which compare individual distributional summaries, Figure~\ref{fig:densities} displays the complete predicted \PMten concentration density functions, providing a more comprehensive assessment of the agreement between MQSR and BSA-FRK.
The two approaches exhibit remarkable agreement in identifying the location of the dominant mode and the distributional shape, indicating highly consistent predictions of \PMten concentrations across municipalities with different levels of regulatory risk.

Taken together, these results indicate that the differences among the three approaches are largely confined to municipalities located at the boundary of the high-risk area, whereas the overall spatial pattern remains essentially unchanged across the different distributional representations.

\section{\authors{Block cross-validation comparison}}
\label{subsection:CV}

\authors{While the previous section focused on qualitative comparisons of the predicted distributions, we now evaluate the out-of-sample predictive performance of the three proposed approaches through a spatial block cross-validation (CV) procedure. By constructing spatially separated training and test sets, this strategy mitigates the effect of spatial autocorrelation, a well-known limitation of conventional CV in spatial regression \citep{roberts2017cross}. Details on the construction of the test folds are provided in Appendix~\ref{section:appendix-b}.}

\authors{We further compare the proposed distributional approaches with a classical geostatistical benchmark based on univariate summary statistics. Specifically, we consider Universal Block Kriging (UBK) \citep{cressie1993statistics} for modeling the probability of exceeding the regulatory \PMten threshold, using the covariates introduced in Section~\ref{section:data}, a spherical variogram, and a nugget effect.}

\authors{We first assess the predictive accuracy of UBK and the three proposed approaches for estimating the exceedance probability \(p\). Performance is evaluated using the Symmetric Mean Absolute Percentage Error (SMAPE), defined as
\[
\text{SMAPE} = 2\frac{|\hat{p} - p|}{\hat{p} + p},
\]
where \(\hat{p}\) and \(p\) denote the predicted and observed exceedance probabilities, respectively.}

\begin{figure}[tb]
    \centering
      \caption{\authors{Block cross-validation results for the competing approaches across the 30 test folds. Left: boxplots of the Symmetric Mean Absolute Percentage Error (SMAPE) for the predicted probability of exceeding the regulatory \PMten threshold of \(50\,\mu g/m^3\). Right: boxplots of the 2-Wasserstein error for the predicted \PMten concentration density functions.}%
      }
    
    \includegraphics[width=0.95\textwidth]{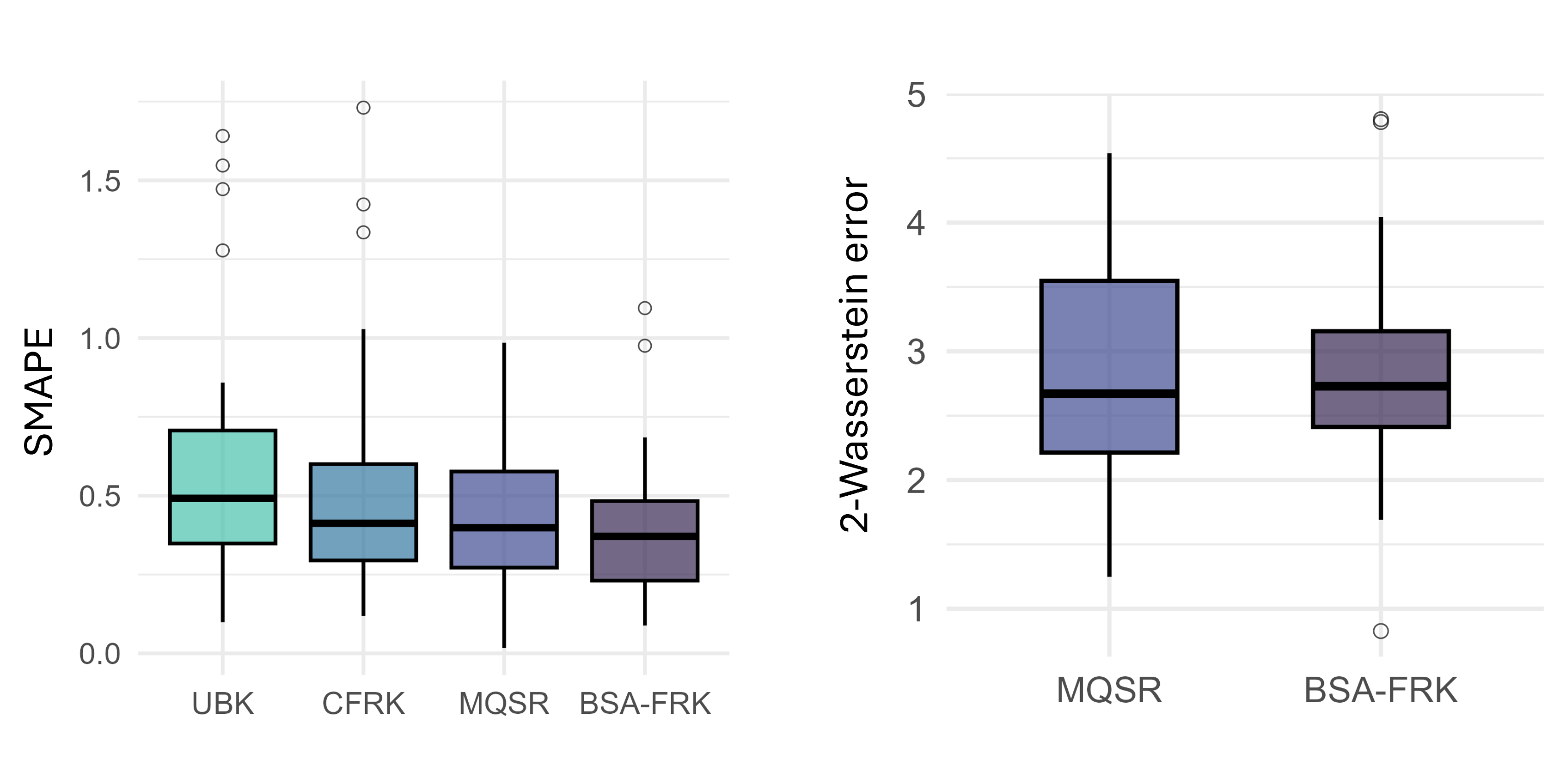} 
    
    \hfill
    \label{fig:validation_results}
\end{figure}

\authors{The results are reported in the left panel of Figure~\ref{fig:validation_results}. Relative to UBK, the three proposed approaches CFRK, MQSR, and BSA-FRK reduce the average SMAPE by 11.5\%, 27.0\%, and 34.5\%, respectively. These improvements are supported by unilateral pairwise Wilcoxon tests. The results show that modeling the full distribution yields a measurable gain in predictive accuracy even when interest is restricted to a single distributional summary. Moreover, the proposed distributional approaches provide a unified framework from which multiple distributional summaries can be derived while preserving their mathematical coherence, avoiding the need to fit separate scalar models for each quantity of interest.}

\begin{figure}[tb]
    \centering
  \caption{\authors{Out-of-sample density estimation of \PMten concentrations for the municipalities of Milano (left) and Sondrio (right). Histograms represent the empirical distribution of the test data, while the solid curves show the probability density functions predicted by Multiple Quantile Spatial Regression (red) and Bayes Space Alignment with Fixed Rank Kriging (green).}}
    \includegraphics[width=\textwidth]{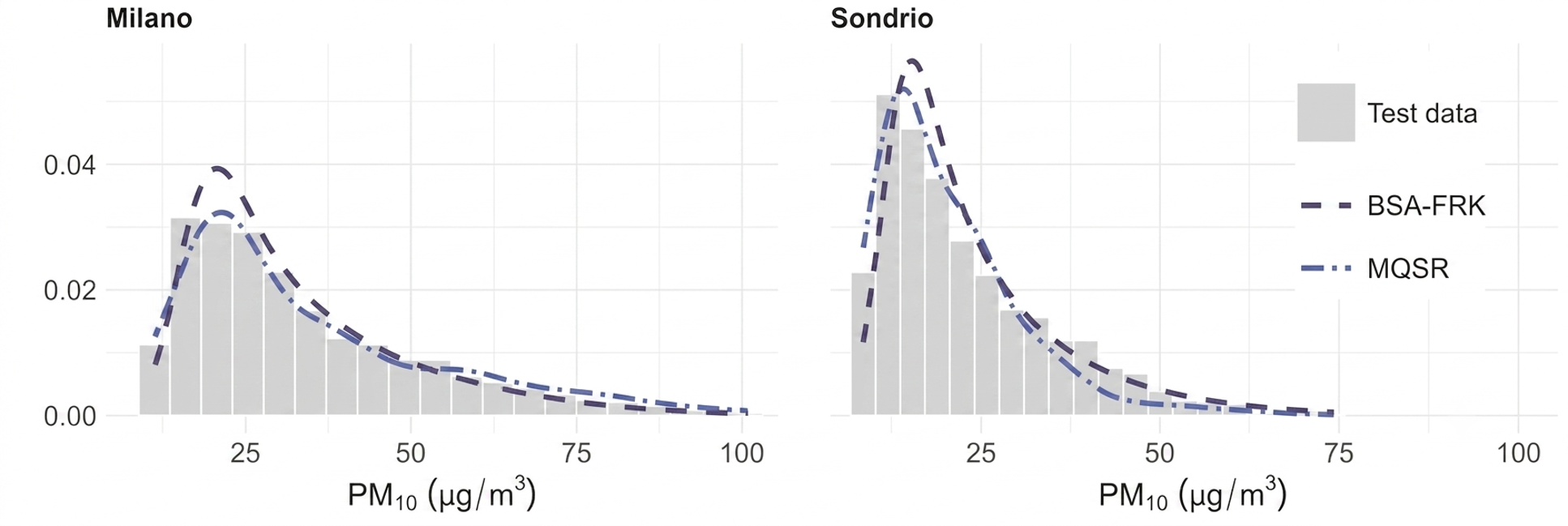}
    \label{fig:validation_results-dens-comuni}
\end{figure}

\authors{
We further assess the predictive accuracy of MQSR and BSA-FRK in reconstructing the full probability density functions. The right panel of Figure~\ref{fig:validation_results} reports the CV prediction errors of the two approaches, measured using the 2-Wasserstein distance \citep{panaretos2019statistical} that, for univariate distributions, is defined as
\[
\sqrt{\int_0^1 \bigg(Q(\alpha) - \hat{Q}(\alpha) \bigg)^2 \, \text{d}\alpha}.
\]
Here, \(Q(\cdot)\) and \(\hat{Q}(\cdot)\) denote the empirical and predicted quantile functions, respectively, and the above integral is evaluated on an equally spaced grid of 50 quantile levels. The two approaches achieve nearly identical median prediction errors, although BSA-FRK exhibits lower variability across the test folds.}

\authors{Finally, Figure~\ref{fig:validation_results-dens-comuni} illustrates the out-of-sample predicted \PMten concentration density functions for the municipalities of Milano (left) and Sondrio (right), together with the corresponding test data. These municipalities represent two contrasting environmental settings in Northern Italy. Milano lies within the Po Valley, where pollution levels are typically high, whereas Sondrio is located in the Alpine region, characterized by generally lower \PMten concentrations and a sparser monitoring network. In both cases, the predicted density functions closely match the histograms of the test data, providing further evidence of the good predictive performance of the two approaches.
}

\section{Discussion and Conclusions}
\label{section:discussions}

This work investigated three spatial approaches for modeling the full distribution of \PMten concentrations across Northern Italy. Although based on different distributional representations, the three approaches yielded highly consistent spatial patterns and improved predictive performance over a classical geostatistical benchmark. \authors{Together, these results show that adopting a distributional perspective can provide tangible advantages for air quality assessment, even when the primary interest lies in a single distributional summary, such as the exceedance probability.}

\authors{Beyond the improvement in predictive accuracy, the proposed approaches provide a unified representation from which multiple distributional summaries can be derived while preserving their mathematical coherence. This avoids fitting separate scalar models for individual quantities of interest and offers a flexible framework for environmental monitoring and regulatory risk assessment.}

\authors{The present work also has several limitations. First, the proposed analyses treat the observations collected at each monitoring station as realizations from a single probability distribution and therefore do not explicitly model the temporal evolution of \PMten concentrations. Extending the proposed framework to spatio-temporal distributional models would allow the investigation of seasonal variability and long-term temporal trends. Second, the analyses focus on a single pollutant. Extending the proposed framework to jointly model the distributions of multiple pollutants represents an interesting direction for future research, although it poses substantial methodological and computational challenges.}

\authors{The encouraging results obtained in this study suggest that explicitly modeling the full distribution represents a promising direction for spatial environmental analyses. While motivated by air quality assessment, the proposed framework is readily applicable to other environmental settings in which the full distribution of the response, rather than a single summary statistic, is object of interest.}\\

%%%%%%%%%%%%%%%%%%%%%%%%%%%%%%%%%%%%%%%%%%%%%%%%%%%%%%%%%%%%%%%%%%%%%%%%%%%

% ringraziamenti (commentati per double-blinded review)
%\begin{comment}
   \backmatter
\bmhead{Acknowledgements}
All the authors acknowledge the project GRINS - Growing Resilient, INclusive and Sustainable (GRINS PE00000018 – CUP D43C22003110001), funded by the European Union -  NextGenerationEU programme. The views and opinions expressed are solely those of the authors and do not necessarily reflect those of the European Union, nor can the European Union be held responsible for them. L.M.\ Sangalli also acknowledges the PRIN2022 project CoEnv - Complex Environmental Data and modelling (CUP2022E3RY23) founded by the European Union - NextGenerationEU program and by the Italian Ministry for University and Research (MUR). 
%\end{comment}

%%%%%%%%%%%%%%%%%%%%%%%%%%%%%%%%%%%%%%%%%%%%%%%%%%%%%%%%%%%%%%%%

\textbf{Conflict of interest:} The authors certify that they have no affiliation with or involvement in any organization or entity with any financial or non-financial interest in the subject of matters or materials discussed in this manuscript.

\bibliography{bibliography.bib}

%%%%%%%%%%%%%%%%%%%%%%%%%%%%%%%%%%%%%%%%%%%%%%%%%%%%%%%%%%%%%%%%%%%%%%%%%%%%%%%%%
\begin{appendices}

\section{Dataset construction}
\label{section:appendix-a}

This appendix describes the procedure used to construct the \PMten dataset analyzed in this study, starting from the air quality measurements provided by the EEA Air Quality Download Service (\url{https://eeadmz1-downloads-webapp.azurewebsites.net/}). For monitoring stations located in Italy, the data distributed by the EEA were originally collected by the corresponding Regional Agencies for Environmental Protection (ARPAs).

The original dataset includes observations at two temporal resolutions, depending on the monitoring station: daily averages, which correspond to the temporal aggregation adopted by the regulatory standards, and hourly averages, which provide higher-resolution measurements that capture short-term pollution patterns associated with the daily cycles of human activity.
As described in Section~\ref{section:data}, the analyses presented in this work are based on geo-referenced time series of daily average \PMten concentrations collected between 2018 and 2022 from 266 monitoring stations across Northern Italy, each uniquely identified by a \texttt{NAT} code. During this period, some sensors were replaced or newly installed at existing monitoring stations, resulting in a total of 325 distinct sensors.

The preprocessing pipeline consisted of three main steps: harmonizing the temporal resolution of the measurements, merging the time series from sensors associated with the same monitoring station, and trimming extreme observations, as described below. All measurements were harmonized to a daily temporal resolution by computing daily averages according to the same criteria adopted by the ARPAs.

\begin{figure}
\centering
\caption{Spatial distribution of the monitoring sensors measuring \PMten concentrations across Northern Italy during the period 2018--2022 (left). Different symbols identify the measurement principles adopted by the sensors: beta radiation, gravimetry, OPC, nephelometry, and TEOM. Top right: number of sensors grouped by measurement principle. Bottom right: partially overlapping time series recorded at station 100114 by two sensors (identified by their Sampling Point IDs) during 2021.}
\label{fig:monitoring-stations-equipments}
\vspace{7pt}
\includegraphics[width=0.9\linewidth]{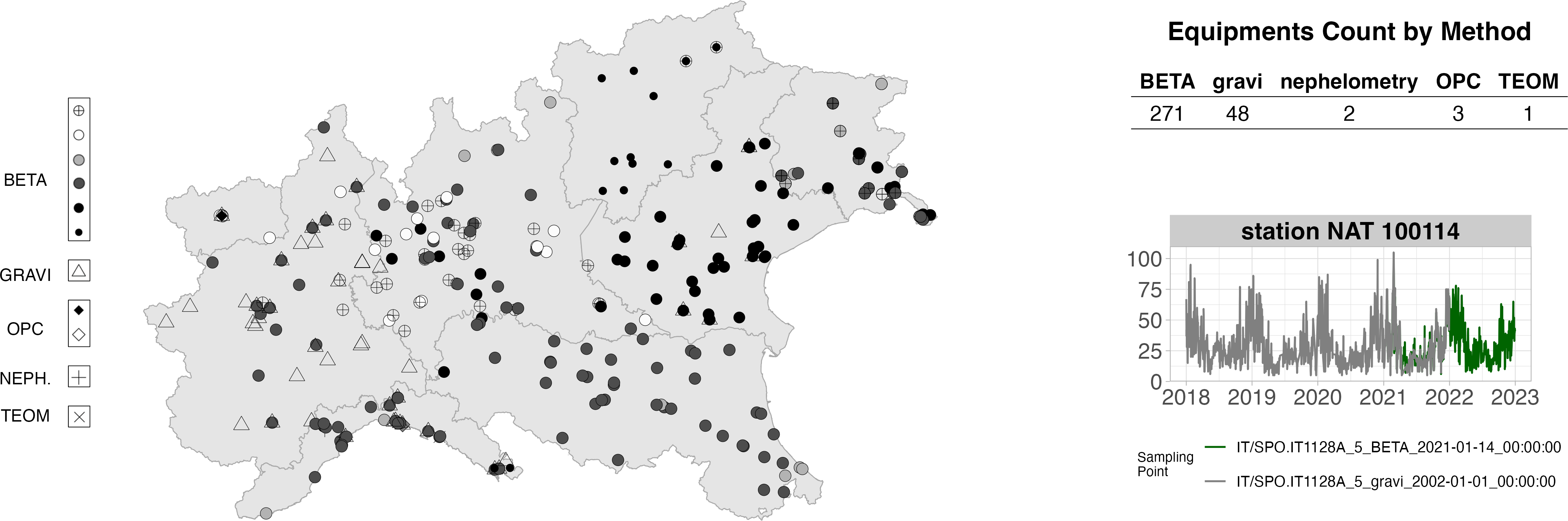}
\end{figure}

Figure~\ref{fig:monitoring-stations-equipments} displays the spatial distribution of the monitoring sensors across Northern Italy. Each sensor is associated with a Sampling Point ID, which provides information on the measurement equipment installed. Different symbols identify the various sensor types, which can be grouped according to their measurement principle: beta radiation, gravimetry, OPC, nephelometry, and TEOM. Detailed information on all sensors, including their technical specifications, is available through the EEA portal (\url{https://discomap.eea.europa.eu/App/AQViewer/index.html?fqn=Airquality_Dissem.b2g.measurements}).

The spatial distribution of the measurement technologies largely reflects the administrative organization of the regional monitoring networks. Nevertheless, some monitoring stations experienced changes in instrumentation over time. During periods in which multiple sensors operated simultaneously at the same monitoring station, we retained only the subset of sensors involving the smallest number of distinct equipment types, thereby reducing variability attributable to instrumentation differences. 
For example, at the station identified by \texttt{NAT} 100114, two partially overlapping time series were collected during 2021 using sensors based on different measurement principles. Following the criterion described above, we retained the observations acquired with the beta radiation sensor to maximize consistency within the resulting time series.

As discussed in Section~\ref{section:data}, the presence of extreme observations represents a common challenge in the analysis of environmental data, particularly air quality measurements \citep{pm10_sources_2}. These observations often arise from exceptional anthropogenic events, such as fireworks, bonfires, and wildfires (see Figure~\ref{fig:monitoring-stations}), or from meteorological conditions and agricultural activities that can produce unusually high or low \PMten concentrations. Examples include storms, strong winds, dust storms, droughts, and manure spreading, all of which may generate transient pollution levels that are not representative of the long-term distribution of the phenomenon.
For example, \citet{vecchi2008impact} showed that metals released during a series of pyrotechnic events in Milan in July 2006 caused \PMten concentrations to increase to nearly four times their typical levels before returning to baseline within approximately four hours.

\begin{figure}
\centering
\caption{Examples of extreme \PMten observations recorded at four monitoring stations. From left to right and top to bottom: elevated concentrations associated with New Year's Eve fireworks (red); a prolonged pollution episode following a severe drought (blue); an isolated peak likely associated with manure spreading in a rural area (green); and a sharp increase associated with a nearby wildfire (brown). Concentrations are measured in $\mu g/m^3$.}
\label{fig:esempi_estremi}
\vspace{7pt}
\includegraphics[width=0.9\linewidth]{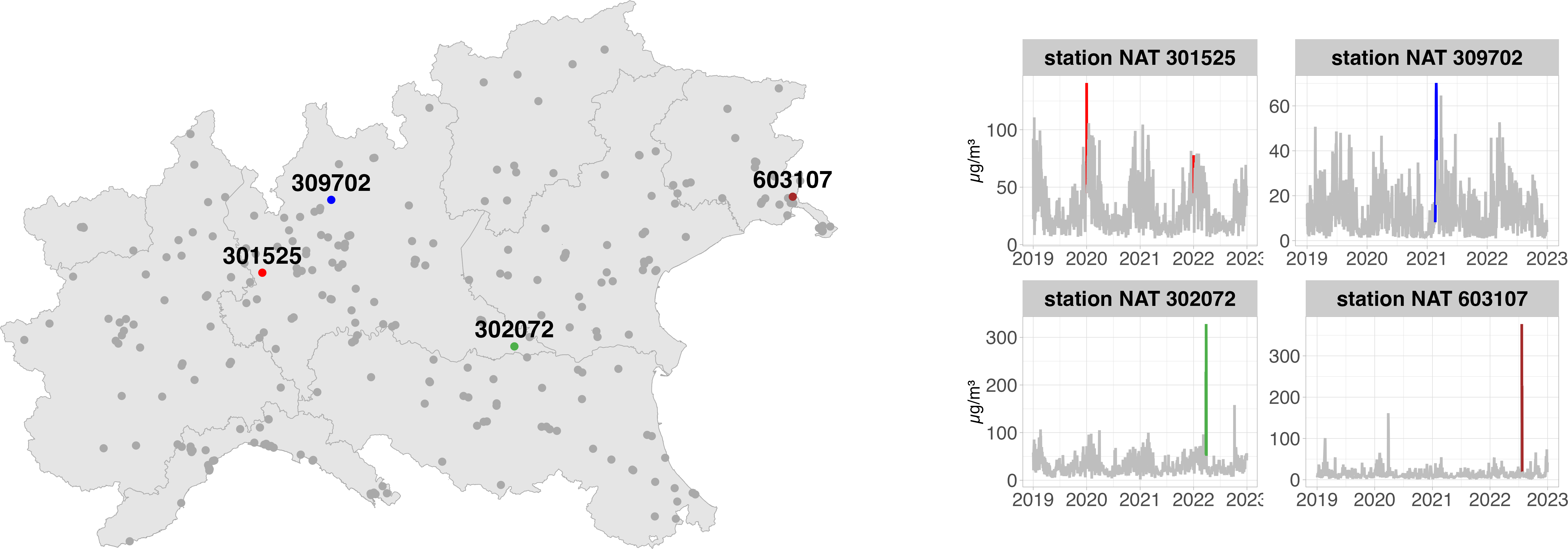}
\end{figure}

Figure~\ref{fig:esempi_estremi} illustrates representative examples of extreme observations identified in the \PMten time series, all characterized by clearly isolated spikes.
The first panel, corresponding to station 301526, shows peaks associated with fireworks, a common source of short-term increases in \PMten concentrations that are particularly evident around New Year's Eve, although similar episodes may occur at other times of the year. Station 301401 (second panel) illustrates an extreme pollution episode caused by exceptional meteorological conditions during the severe drought affecting the Po Valley in February 2021 (\url{https://www.lombardianotizie.online/antismog-misure-primo-livello/},
\url{https://www.assolombarda.it/servizi/mobilita-e-trasporti/informazioni/regione-lombardia-attivate-le-misure-temporanee-di-limitazione-alla-circolazione}). Similar anomalies were observed at several other monitoring stations during the same period.
The third panel, corresponding to station 302072, shows an isolated peak likely associated with manure spreading activities near the monitoring station. Finally, station 603107 (fourth panel) exhibits a sharp increase in late July 2022, corresponding to a wildfire that occurred close to the monitoring site and lasted approximately two days (\url{https://www.arpa.fvg.it/temi/temi/aria/news/incendio-sul-carso-monfalconese-lisert-aggiornamento-misure/}). These examples illustrate the heterogeneous origin of extreme observations and motivate the trimming procedure adopted to characterize the structural distribution of \PMten concentrations.

\begin{figure}
\centering
\caption{Example of the trimming procedure applied to the \PMten time series recorded at station 100167. The dashed lines represent the estimated station-specific quantile thresholds, \(Q_1\) and \(Q_{99}\). Observations exceeding \(Q_{99}\) (dark red) or falling below \(Q_1\) (steel blue) were identified as outliers and removed from the analysis.}
\label{fig:removed_values}
\vspace{7pt}
\includegraphics[width=0.6\linewidth]{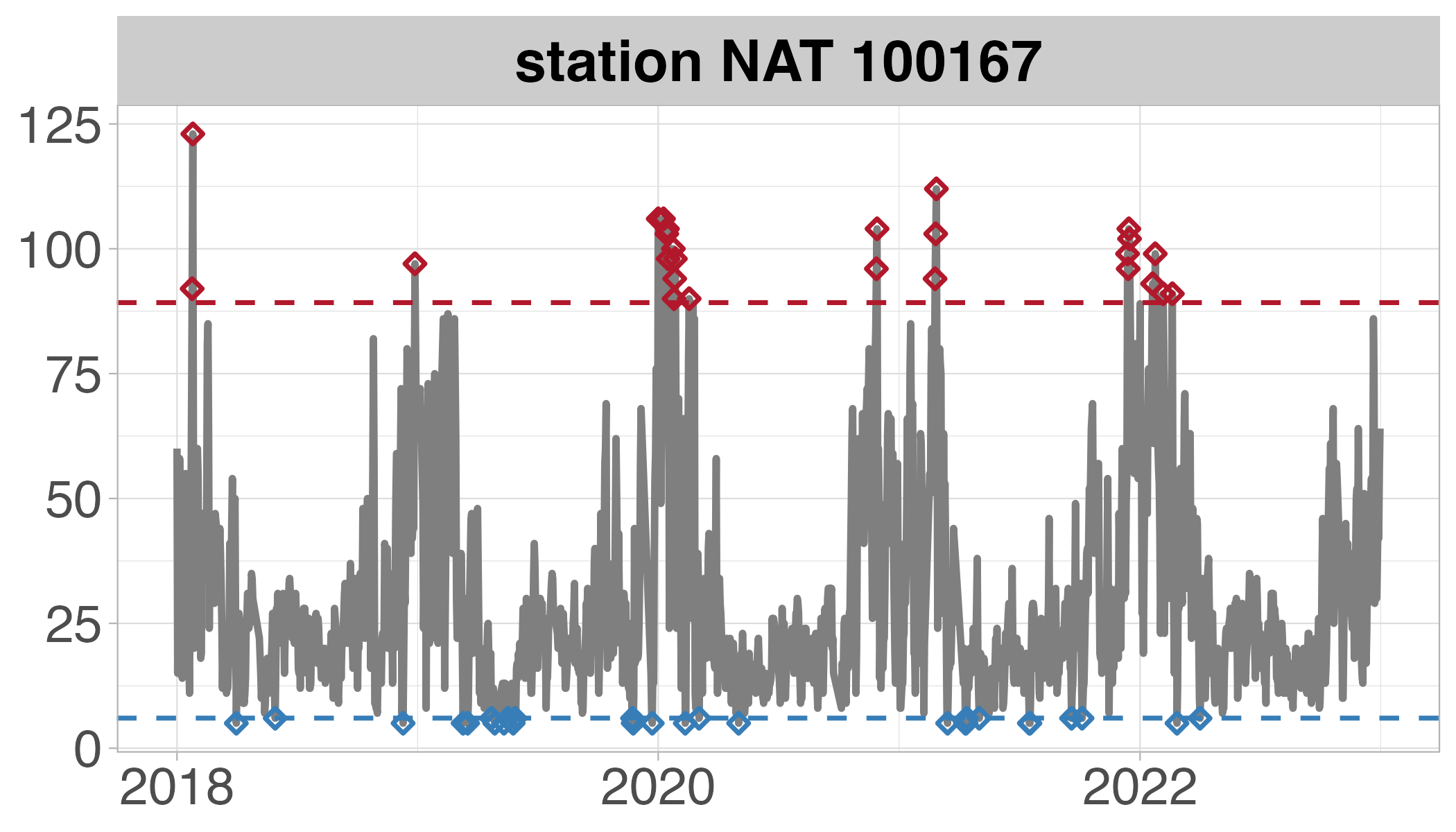}
\end{figure}

As discussed in Section~\ref{section:data}, sporadic events may not be representative of the underlying spatial distribution of \PMten concentrations. We therefore estimate the 0.01 and 0.99 spatial quantile fields using the method proposed by \citet{castiglione_2025}, and use the resulting station-specific thresholds to remove observations below \(Q_1\) and above \(Q_{99}\). Figure~\ref{fig:removed_values} illustrates this procedure for a representative monitoring station, showing the estimated thresholds together with the observations identified as outliers.

\section{Cross-Validation Procedure}
\label{section:appendix-b}

\begin{figure}[h]
    \centering
       \caption{\authors{Spatial distribution of the $30$ cross-validation folds across Northern Italy. The study region is partitioned into a regular \(10 \times 10\) grid of rectangular blocks, with municipalities assigned to each block according to centroid coordinates. Cross-validation folds, containing different blocks, are randomly chosen over the domain. The municipalities highlighted in the picture are those with at least one monitoring stations (black dots), and are colored according to the fold the belong to.}}
    \includegraphics[width=\textwidth]{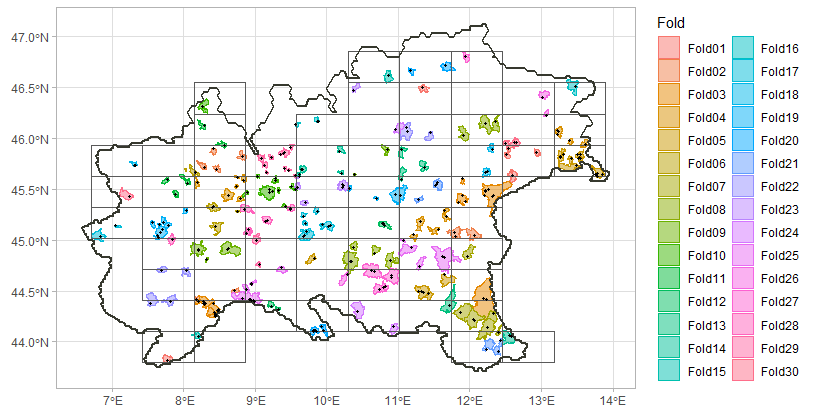} 
    \label{fig:cv_folds}
\end{figure}

\authors{This appendix details the Cross-Validation (CV) procedure introduced in Section \ref{subsection:CV}, which employs the \texttt{spatial\_block\_cv} function from the R package \texttt{spatialsample} \citep{mahoney2023assessing}. 
The method first overlays a $10 \times 10$ regular grid across the Northern Italy study region. The algorithm assigns each municipality containing at least one monitoring station to a specific grid element based on the municipal centroid coordinates. This step partitions the municipalities with at least one observation into subsets corresponding to these grid blocks. Then these subsets are randomly assigned to $30$ distinct spatial folds, meaning each fold combines municipalities from multiple grid blocks, as shown in Figure \ref{fig:cv_folds}.}

\end{appendices}

\end{document}